\documentclass[apj,twocolumn,twocolappendix,numberedappendix]{openjournal}
\usepackage{amsmath}
\usepackage{booktabs}
\usepackage{multirow}
\usepackage{color}
\usepackage{soul}
\usepackage{threeparttable}
\usepackage{float}
\usepackage{graphicx}
\usepackage{CJK}
\usepackage{xspace}
\usepackage{afterpage}
\usepackage{placeins}
\usepackage{amssymb}
\usepackage{enumitem}
\usepackage[breaklinks,colorlinks,citecolor=blue,urlcolor=blue,linkcolor=blue,filecolor=blue]{hyperref}

\makeatletter
  \patchcmd{\NAT@citex}
    {\@citea\NAT@hyper@{%
      \NAT@nmfmt{\NAT@nm}%
      \hyper@natlinkbreak{\NAT@aysep\NAT@spacechar}{\@citeb\@extra@b@citeb}%
      \NAT@date}}
    {\@citea\NAT@nmfmt{\NAT@nm}%
    \NAT@aysep\NAT@spacechar\NAT@hyper@{\NAT@date}}{}{}
  \patchcmd{\NAT@citex}
    {\@citea\NAT@hyper@{%
      \NAT@nmfmt{\NAT@nm}%
      \hyper@natlinkbreak{\NAT@spacechar\NAT@@open\if*#1*\else#1\NAT@spacechar\fi}%
        {\@citeb\@extra@b@citeb}%
      \NAT@date}}
    {\@citea\NAT@nmfmt{\NAT@nm}%
    \NAT@spacechar\NAT@@open\if*#1*\else#1\NAT@spacechar\fi\NAT@hyper@{\NAT@date}}
    {}{}
\makeatother

\shorttitle{LRDs as BH*s (with winds)}
\shortauthors{Naidu et al.}

\usepackage[export]{adjustbox}

\newcommand{\orcidauthor}[3]{\author{\href{http://orcid.org/#1}{#2$^{#3}$}}}

\begin{document}

\title{\vspace{-1cm} Little Red Dots as Intermediate Mass, Super-Eddington Engines:\\Insights from Type II\MakeLowercase{n} Supernovae and The 1837-1856 Great Eruption of $\eta$ Carinae \vspace{-1.75cm}}

\orcidauthor{0000-0003-3997-5705}{Rohan P. Naidu}{1, 2, *, \dagger}
\orcidauthor{0000-0003-2871-127X}{Jorryt Matthee}{3}
\orcidauthor{0000-0002-2380-9801}{Anna de Graaff}{4, 5, \ddagger}
\orcidauthor{0000-0001-5586-6950}{Alberto Torralba}{3}
\orcidauthor{0000-0002-5221-7557}{Chris Ashall}{2}
\orcidauthor{0000-0003-1561-3814}{Harley Katz}{6, 7}
\orcidauthor{0000-0002-0302-2577}{John Chisholm}{8, 9}
\orcidauthor{0000-0003-2680-005X}{Gabriel Brammer}{10, 11}
\orcidauthor{0000-0003-0599-8407}{Luc Dessart}{12, 13}
\orcidauthor{0000-0003-2895-6218}{Anna-Christina Eilers}{1}
\orcidauthor{0000-0002-4684-9005}{Raphael E. Hviding}{5}
\orcidauthor{0000-0002-6230-0151}{David~O.~Jones}{14}
\orcidauthor{0000-0002-5588-9156}{Vasily Kokorev}{8, 9}
\orcidauthor{0000-0001-6755-1315}{Joel Leja}{15, 16, 17}
\orcidauthor{0000-0003-2488-4667}{Hanpu Liu}{18}
\orcidauthor{0009-0002-8965-1303}{Zhaoran Liu}{1}
\orcidauthor{0000-0003-1927-4397}{Devesh Nandal}{4}
\orcidauthor{0000-0001-5851-6649}{Pascal A.\ Oesch}{19, 10, 11}
\orcidauthor{0000-0003-4175-4960}{Conor L. Ransome}{20}
\orcidauthor{0000-0003-3769-9559}{Robert A. Simcoe}{1}
\orcidauthor{0009-0007-3791-7890}{Wendy Q. Sun}{1}
\orcidauthor{0000-0001-8928-4465}{Andrea Weibel}{19}
\orcidauthor{0000-0003-1207-5344}{Mengyuan Xiao}{19}

\affiliation{$^{1}$ MIT Kavli Institute for Astrophysics and Space Research, 70 Vassar Street, Cambridge, MA 02139, USA}
\affiliation{$^{2}$ Institute for Astronomy, University of Hawai‘i, 2680 Woodlawn Drive, Honolulu, HI 96822, USA}
\affiliation{$^{3}$ Institute of Science and Technology Austria (ISTA), Am Campus 1, 3400 Klosterneuburg, Austria}
\affiliation{$^{4}$ Center for Astrophysics, Harvard \& Smithsonian, 60 Garden St, Cambridge, MA 02138, USA}
\affiliation{$^{5}$ Max-Planck-Institut f\"ur Astronomie, K\"onigstuhl 17, D-69117 Heidelberg, Germany}
\affiliation{$^{6}$ Department of Astronomy \& Astrophysics, University of Chicago, 5640 S Ellis Avenue, Chicago, IL 60637, USA}
\affiliation{$^{7}$ Kavli Institute for Cosmological Physics, University of Chicago, Chicago IL 60637, USA}
\affiliation{$^{8}$ Department of Astronomy, The University of Texas at Austin, Austin, TX, USA}
\affiliation{$^{9}$ Cosmic Frontier Center, The University of Texas at Austin, Austin, TX 78712, USA}
\affiliation{$^{10}$ Cosmic Dawn Center (DAWN), Copenhagen, Denmark}
\affiliation{$^{11}$ Niels Bohr Institute, University of Copenhagen, Jagtvej 128, K{\o}benhavn N, DK-2200, Denmark}
\affiliation{$^{12}$ French-Chilean Laboratory for Astronomy, IRL 3386, CNRS and Instituto de Astrofísica, Pontificia Universidad Católica de Chile, Casilla 306, Santiago, Chile}
\affiliation{$^{13}$ Institut d'Astrophysique de Paris, CNRS-Sorbonne Universit\'e, 98 bis boulevard Arago, F-75014 Paris, France}
\affiliation{$^{14}$ Institute for Astronomy, University of Hawai‘i, 640 N Aohoku Place, Hilo, HI 96720, USA}
\affiliation{$^{15}$ Department of Astronomy \& Astrophysics, The Pennsylvania State University, University Park, PA 16802, USA}
\affiliation{$^{16}$ Institute for Gravitation and the Cosmos, The Pennsylvania State University, University Park, PA 16802, USA}
\affiliation{$^{17}$ Institute for Computational \& Data Sciences, The Pennsylvania State University, University Park, PA 16802, USA}
\affiliation{$^{18}$ Department of Astrophysical Sciences, Princeton University, Princeton, NJ 08544, USA}
\affiliation{$^{19}$ Department of Astronomy, University of Geneva, Chemin Pegasi 51, 1290 Versoix, Switzerland}
\affiliation{$^{20}$ Steward Observatory, University of Arizona, 933 North Cherry Avenue, Tucson, AZ 85721-0065, USA}

\thanks{$^*$E-mail: \href{mailto:rohan.naidu@hawaii.edu}{rohan.naidu@hawaii.edu}}
\thanks{$\dagger$ NASA Hubble Fellow, Pappalardo Fellow}
\thanks{$\ddagger$ Clay Fellow}

\begin{abstract}
    JWST's Little Red Dots (LRDs) display a unique constellation of features that do not occur simultaneously in any other class of galaxies or AGN. Here we observe that many of these features find parallels in the 19th century Great Eruption (GE) of $\eta$ Carinae and a sub-class of supernovae (Type IIn). Drawing on these stellar phenomena -- outflows trapped by dense circumstellar gas envelopes -- we sketch a possible scenario for LRDs. Outflows from the central engine produce an enshrouding envelope of gas that may be thought of as a slow wind. This dense wind and its enormous extent produce an opacity so high that a pseudo-photosphere forms within the wind, obscuring the central engine and manifesting as a blackbody-like continuum. Radiation from the buried engine powers the system. The engine may also launch fast winds that crash into the existing envelope to generate shocks. Lines form within the wind above the photosphere -- electron scattering and absorption in the clumpy (ionized + neutral) medium account for broad wings and P-Cygni cores. A key implication is that inferences of ``overmassive black holes" may be interpreting this wind-like physics as a virial broad-line region. We propose an escape velocity argument to constrain the mass of the engine, which yields $M<10^{5} M_\odot$ for the typical LRD. The lack of variability and low surface gravity of the photosphere provide further support for intermediate mass ($M\approx10^{3-6} M_\odot$), but very luminous super-Eddington ($L_{\rm{bol}}/L_{\rm{edd}}\gtrsim5$) systems, harboring a supermassive star or intermediate mass black hole. Paralleling the evolution of IIn SNe, dust production in the envelope may mark the beginnings of classical AGN. This paper explores a possible self-consistent explanation for the entire life-cycle of LRDs, from their enshrouding in dense gas to their fates as seeds of massive black holes.
\end{abstract}

\section{Introduction}
\label{sec:intro}

In 1843, the massive star $\eta$ Carinae brightened to such a degree ($\approx5$ mags brighter than today) that it became the second brightest star in the entire sky \citep[e.g.,][]{Smith11}. For almost two decades, $\eta$ Car was so prominent that the aboriginal Boorong people of Australia adjusted their mythology of the Carina constellation -- the newly luminous and newly reddened $\eta$ Car was accorded the status of the wife of the other luminous star, Canopus, and the mother of the fainter surrounding stars \citep[e.g.,][]{Hamacher10}. Light echoes of this ``Great Eruption" (GE) reflected from surrounding dust clouds were detected in the 2010s, allowing us to examine this remarkable event with modern spectrographs \citep[e.g.,][]{Rest12, Prieto14, Smith18}.

The GE is the most well-known example of a class of objects that we will describe as ``enshrouded eruptions". In the decades preceding the GE, $\eta$ Carinae shed $\approx10-30\%$ of its mass (present-day mass of $\approx100 M_{\rm{\odot}}$) in mini-eruptions, episodes now recognized to be common among massive stars \citep[e.g.,][]{Smith14, Morris17, Smith18}. The result was a dense, slowly outflowing gas envelope enclosing the star. And then in 1837, $\eta$ Car, which at that time was likely in a triple star system (all enclosed within the envelope), merged with one of its companions \citep[e.g.,][]{Zwart16,Hirai21}. This unleashed an injection of energy, which under normal circumstances may have culminated in a short-lived transient. But instead, the ``hot wind" from this merger crashed into the surrounding ``cold wind", setting off a cascade of shocks that slowly worked their way through the dense circumstellar medium \citep[CSM; e.g.,][]{Smith18}. The highly efficient conversion of kinetic energy of the explosive hot wind into radiative energy -- $>50\%$ efficiency, compared to $\lesssim1\%$ in typical Type II SNe \citep{Dessart16, Smith17} --  is what accounts for the extraordinary super-Eddington luminosity of the GE --  $L_{\rm{bol}}/L_{\rm{edd}}>5$ sustained over two decades \citep[e.g.,][]{Rest12}.

Shorter-lived, but spectrally similar, Type IIn SNe constitute the best studied class of enshrouded eruptions \citep[e.g.,][]{ Schlegel90}. In fact, Type IIn SNe are defined by the fact that they are enshrouded -- what is common across IIn SNe is not a kind of central engine, but that they are all core-collapse explosions that go off within a dense circumstellar medium \citep[e.g.,][]{Smith17}. The envelope confines the typically extremely broad lines of SNe ($\gtrsim10,000$ km s$^{-1}$) into narrow cores (hence the ``n" in IIn). As a result of this unique enshrouded configuration, IIn spectra are highly distinctive. For example, the emission lines show a combination of broad wings from electron scattering in ionized gas alongside deep, P-Cygni-like absorption from neutral gas \citep[e.g.,][]{Fassia01, Dessart09, Kiewe12, Taadia13, Kankare25}. 

The blackbody-like continuum seen across the majority of IIn lifetimes (typically $\approx100$ days at $>50\%$ from peak; \citealt{Hiramatsu24}) is reminiscent of the stellar spectrum of a low surface gravity hypergiant with deep absorption features \citep[e.g.,][]{Dessart16}. However, we are not seeing a stellar photosphere in hydrostatic equilibrium, but instead a ``pseudo-photosphere" or ``wind photosphere" or ``recombination photosphere" \citep[e.g.,][]{Leitherer85, Davidson87, Humphreys94, Dessart11, Owocki16}. This is not the actual surface of a star, but the effective radiating surface in the wind where the optical depth reaches unity. The GE, for example, displays a $\approx5000$ K pseudo-photospheric continuum, while the $\approx40,000$ K surface of $\eta$ Car at its center is completely obscured by the enshrouding gas \citep[e.g.,][]{Hillier01, Rest12, Prieto14, Smith18}. The gas is not in local thermodynamic equilibrium (LTE), but the source function approaches a Planck function. In detail, at fixed apparent $T_{\rm{eff}}$, the key difference from a true stellar photosphere is that the wind photosphere spans a far larger extent accompanied by  a continuous temperature and density gradient instead of the sharp cutoff \citep[e.g.,][]{Davidson87, Dessart16, Davidson16}. Contrast the thin, skin-like photosphere of a supergiant star (a few $\%$ of the stellar radius) with the spectral formation region in a IIn that can span 10s to 100s of au \citep[e.g.,][]{Dessart16}.

This brings us to JWST's enigmatic Little Red Dots (LRDs; \citealt[][]{Matthee24}). The basic nature of their central engines remains debated. Do they host black holes or supermassive stars? \citep[e.g.,][]{Nandal25, Chisholm26, Martins26, Zwick26} Are they the elusive seeds of massive black holes ($\lesssim10^{5} M_{\rm{\odot}}$; e.g., \citealt{Greene25, Umeda25, Yanagisawa26, Gentile26}) or are they instead already so massive that in some cases they come close to outweighing their host galaxies ($\gtrsim10^{7} M_{\rm{\odot}}$; e.g., \citealt{Maiolino25, Juodzbalis25direct, Jones25, Gupta26})? 

The key to the puzzle may lie at the very beginning of the LRD story. What we now understand to be $z\approx0$ analogs of LRDs \citep[][]{Lin25, Ji25lol} were first discovered by \citet{Izotov07}, 15 years prior to JWST's launch. These sources were found in a search for broad-line objects among metal-poor SDSS galaxies. Coming across these objects, \citet{Izotov07} classified them under an ambiguous category: ``IIn/AGN?". The spectra had some of the apparent signs of AGN (e.g., broad Balmer lines), but at the same time were indistinguishable from Type IIn SNe. Follow-up spectroscopy \citep[][]{Izotov08} showed the sources to be invariant, thereby ruling out IIn. Interestingly, IIn SNe themselves first appeared in the literature as ``Seyfert I SNe" \citep[][]{Filippenko89}: ``\textit{Had either of these SNe occurred in the nucleus of a normal galaxy, the nucleus would probably have been classified as a Type I Seyfert if the only data available were low-resolution optical spectra}". Furthermore, \citet[][]{Filippenko89} used single star spectra to interpret SN1987F and SN1988I. In other words, ever since their discovery, IIn SNe have been blurring the lines between black holes and stars.

\begin{figure*}
    \centering
\includegraphics[width=0.7\linewidth]{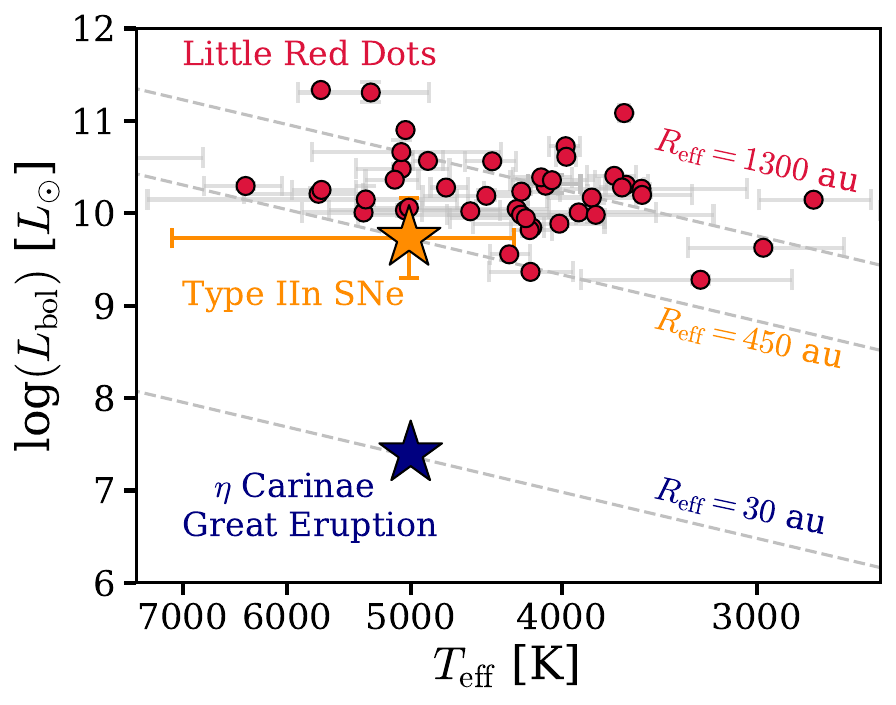}
    \caption{\textbf{LRDs, Type IIn SNe, and the GE span 4 dex in bolometric luminosity ($L_{\rm{bol}}$) but cluster in effective temperature ($T_{\rm{eff}}$)}. The GE (blue star) displays a blackbody-like continuum of $T_{\rm{eff}}\approx5000-6000$ K \citep[][]{Rest12, Smith18} similar to the $\approx500$ Type IIn SNe (dark orange) compiled in \citet[][]{Hiramatsu24} and LRDs (red dots) from \citet[][]{degraaff25pop}. The \citet[][]{Hiramatsu24} IIn SNe compilation is represented at peak luminosity by adopting a flat bolometric correction of $L_{\rm{bol}}/L_{\rm{opt}}=1.9$. The dashed gray lines depict $L_{\rm{bol}}=4\pi R_{\rm{eff}}^{2} \sigma_{\rm{SB}} T_{\rm{eff}}^{4}$ at fixed $R_{\rm{eff}}$. The clustering in $T_{\rm{eff}}$ occurs at the temperatures where hydrogen recombination produces a steep opacity gradient, self-regulating the pseudo-photospheres (or ``recombination photospheres") of these objects \citep[e.g.,][]{Davidson87, Dessart11, Owocki16, Liu25BB, Kido25}. 
    }
\label{fig:hrd}
\end{figure*}

\begin{figure}
    \centering
\includegraphics[width=0.9\linewidth]{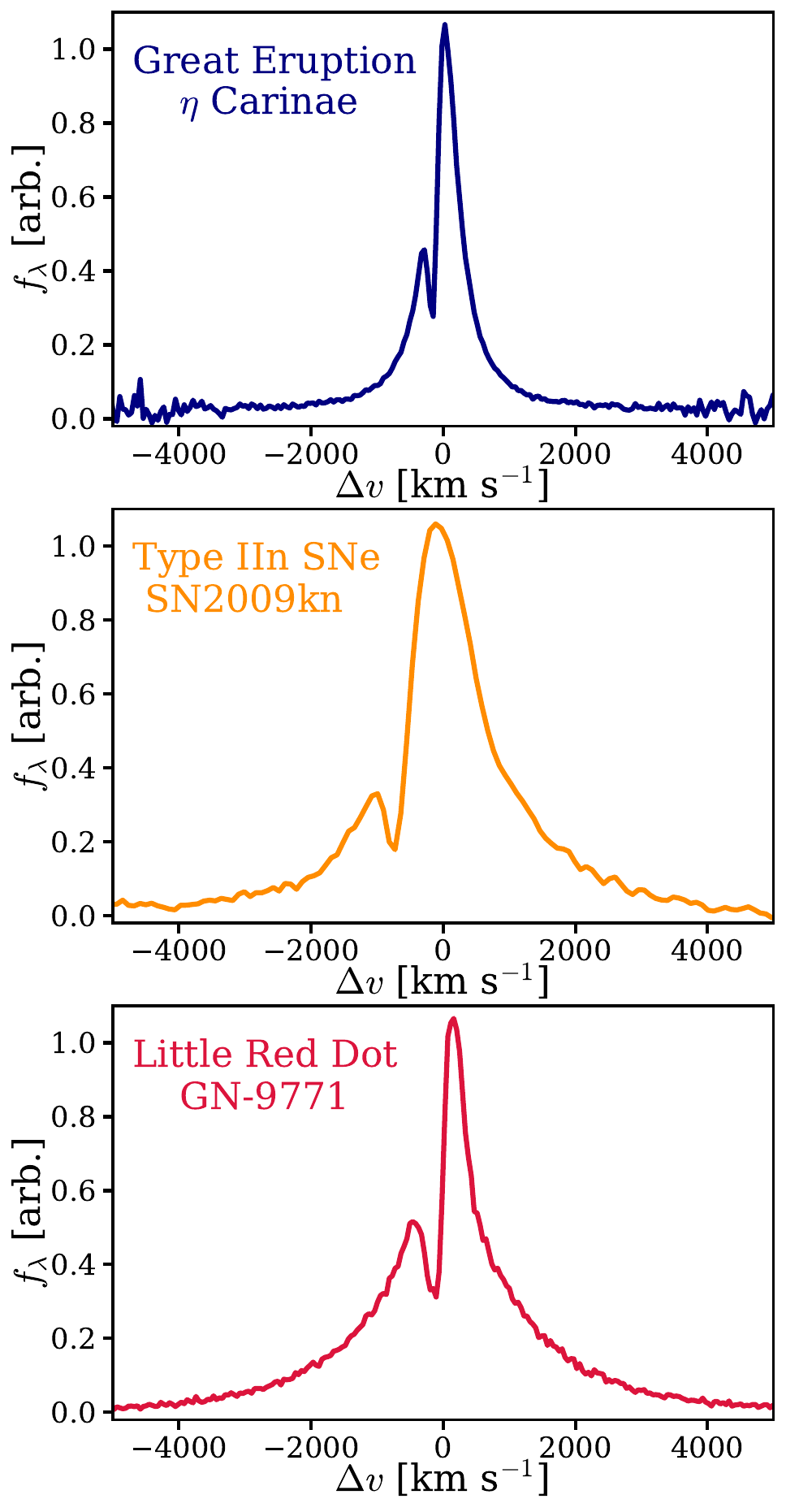}
    \caption{\textbf{The remarkable similarity in H$\alpha$ line profiles of the GE, Type IIn SNe, and LRDs}. \textbf{Top:} In the CSM interaction phase of the GE, the line is characterized by electron scattered wings and blue-shifted Balmer absorption. Profile reconstructed from parameters reported in \citet{Smith18}, matched in resolution and noise properties to the JWST/NIRSpec LRD spectrum in the bottom panel. \textbf{Center:} Type IIn SNe (example from \citealt{Kankare12}; X-SHOOTER spectrum with $R\approx8800$) show very similar line profiles for large periods in their evolution -- narrow lines relative to typical Type II SNe (hence the ``n" in Type IIn), Balmer absorption, and electron scattering wings. Indeed, the GE has been modeled as a ``scaled down" version of such SNe \citep[][]{Smith13}. \textbf{Bottom:} LRDs show similar line profiles (represented here by GN-9771; \citealt{Torralba25IFU}), which in this work we interpret by drawing on the relatively well-understood physics of enshrouded eruptions.
    }
\label{fig:lineprofiles}
\end{figure}

In this paper we propose that the remarkable similarities across LRDs, IIn SNe, and the GE are far from a mere coincidence. \citet{degraaff25} and \citet{Naidu25BHstar} marked a turning point in the LRD story by observing that LRD physics was better understood not in terms of traditional galaxies and AGN, but by drawing analogies to stellar astrophysics. That is, the central engines of LRDs may be thought of as ``black hole stars" (BH*s) -- black holes embedded within dense gas envelopes such that they radiate in a manner reminiscent of stellar phenomena \citep[e.g.,][]{Kido25, Liu25BB, IM25, Inayoshi25coevol, Begelman25}. Indeed, features classically associated with stars are seen among LRDs -- A-star-like Balmer breaks \citep[e.g.,][]{Furtak24, Labbe24, Wang24evolved}, M-dwarf-like water absorption \citep[e.g.,][]{degraaff25pop, Wang26}, photospheric blackbody-like continua \citep[e.g.,][]{degraaff25pop, Umeda25, Sun26, Torralba26panbhstar, Gentile26}, P Cygni line profiles \citep[e.g.,][]{NM24, Juodzbalis24rosetta, Wang24, Loiacono25,Kokorev25glimpsed}, Cepheid-like pulsation (pending spectroscopic confirmation; \citealt[][]{Zhang25cepheid, Zhang26H0, Cantiello25}). The toolkit, language, and analogies to stellar astrophysics have the potential for rich insight into the workings of LRDs -- e.g., the Hertzsprung-Russell (H-R) diagram \citep[e.g.,][]{degraaff25pop, Umeda25, Lin26}, the Hayashi track \citep[e.g.,][]{Kido25, Liu25BB, Chen26}, ``stellar" atmosphere models \citep[][]{Liu26TLUSTY, Santarelli25}, ``stellar" evolutionary grids \citep[][]{Roman-Garza26}, ``proto-star" magnetic fields \citep[][]{Takasao26}, ``stellar" wind models \citep[][]{Chisholm26, Martins26}. Here we hope to extend our understanding of BH*s, drawing new sets of parallels between LRDs and stellar phenomena.

We will first gather the numerous empirical similarities between LRDs and enshrouded eruptions (represented by the GE and IIn SNe\footnote{Many of the LRD features discussed in this paper find echoes among a variety of other stellar phenomena governed by similar pseudo-photospheric physics and high opacity winds: e.g., Luminous Red Novae (LRN), Luminous Blue Variables (LBVs), Type II-P SNe, and the whole zoo of interacting/wind transients \citep[e.g.,][]{Piro20}. We focus here on IIn SNe and the Great Eruption which bear the most striking empirical similarities to LRDs.}) in \S\ref{sec:empirical}. Drawing inspiration from these relatively well-understood stellar phenomena, we will draw insights into the physics of LRDs (\S\ref{sec:physics}) -- as a preview, consider \citet{Davidson16} on giant eruptions of massive stars: \textit{``When we see moderate Thomson-scattering wings on the Balmer emission lines, with a visual continuum slope like $T\approx$7000 to 15000~K, then we’re probably looking at a super-Eddington flow"}. With the insights from the stellar phenomena in mind, in \S\ref{sec:bhmasses} we will outline and pilot arguments for constraining the masses of LRD central engines. We end by discussing broader implications of our proposed picture (\S\ref{sec:discussion}) -- LRD central engines are consistent with being supermassive stars (SMS) and/or intermediate mass black holes (IMBHs), representing the long-sought super-Eddington seeds of present day massive black holes. We adopt a flat $\Lambda$CDM cosmology with parameters as per \citet[][]{Planck18}. For summary statistics, we report medians with uncertainties on the median from bootstrapping (16$^{\rm{th}}$ and 84$^{\rm{th}}$ percentiles).

\section{Empirical Similarities between Little Red Dots and Enshrouded Eruptions}
\label{sec:empirical}

\begin{figure*}
    \centering
\includegraphics[width=0.9\linewidth]{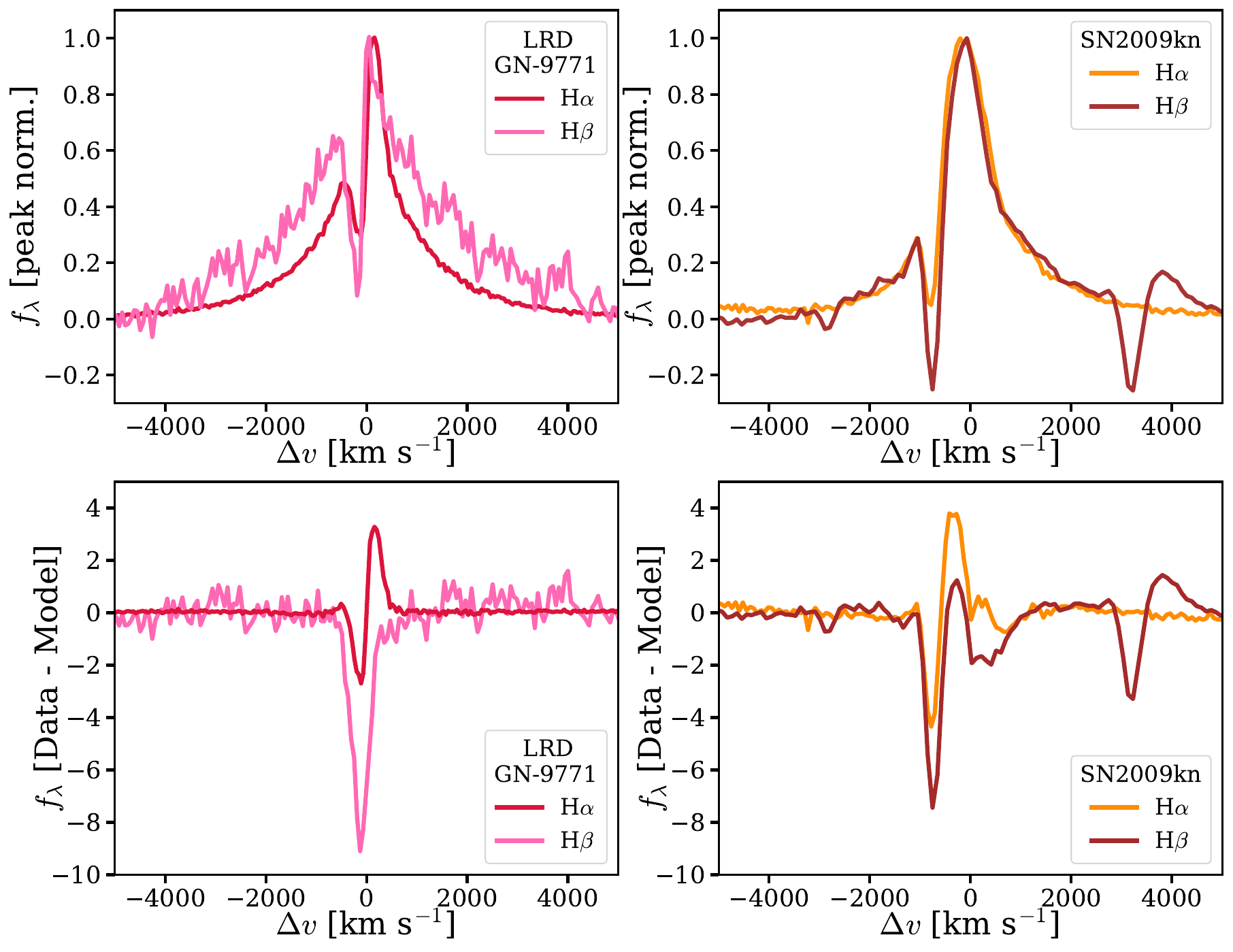}
    \caption{\textbf{Subtle trends in Balmer series line profiles are shared between LRDs and IIn SNe}. \textbf{Top:} Peak-normalized, continuum-subtracted Balmer lines are shown for GN-9771 \citep[][]{Torralba25IFU} and SN2009kn \citep[][]{Kankare12}. \textbf{Bottom:} Following \citet[][]{Matthee26}, a model for the exponential wings (and host galaxy in the case of the LRD) is subtracted from the data. This leaves behind a P-Cygni-like profile, with the trough of H$\beta$ dipping lower than H$\alpha$. This has been successfully modeled in IIn SNe \citep[e.g.,][]{Dessart09} as arising in a partially ionized medium where H$\beta$ forms deeper than H$\alpha$. The deeper forming line sees a larger absorbing column of \ion{H}{1} with $n=2$ level atoms. Other trends in the velocity offsets as well as electron scattering wings across the Balmer series are shared between IIn SNe and LRDs, suggesting very similar radiative transfer processes \citep[e.g.,][]{Dessart09,Brazzini25, Matthee26}. Note that the feature at $\approx3000$ km s$^{-1}$ in the SN2009kn spectrum is due to \ion{Fe}{2} (see Fig. \ref{fig:fullspec}).
    }
\label{fig:hasubtle}
\end{figure*}

\begin{figure}
    \centering
\includegraphics[width=0.9\linewidth]{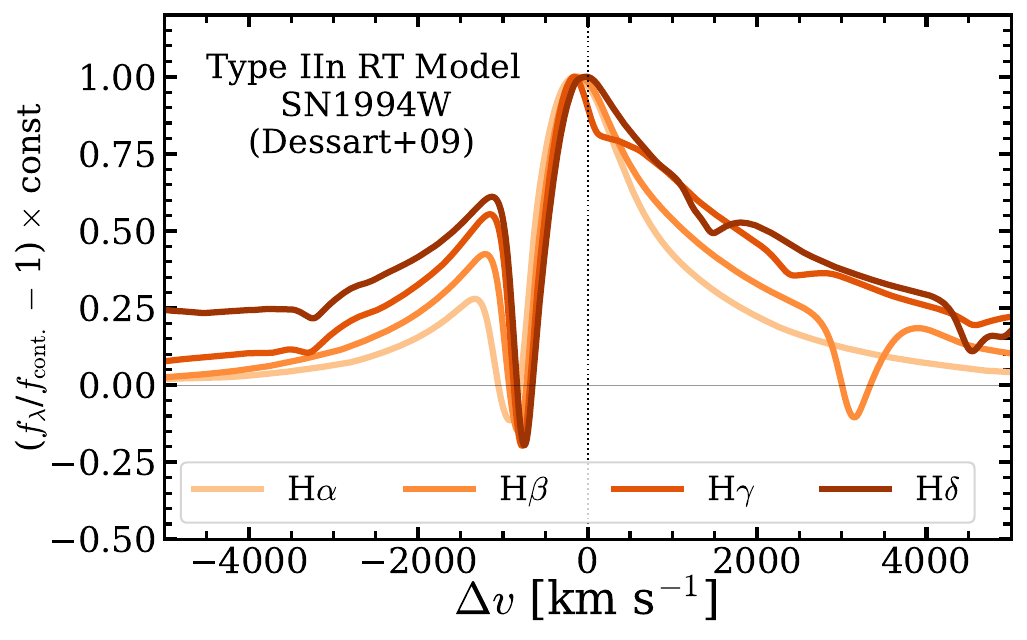}
    \caption{\textbf{Electron scattering explains broad lines in IIn SNe (and LRDs)}. Models from \citet[][]{Dessart09} designed to reproduce SN1994W are shown for illustration. The degree of electron scattering increases down the Balmer series. This is exactly what is observed in LRDs, where H$\beta$ consistently shows broader exponential wings than H$\alpha$ \citep[see Fig. \ref{fig:hasubtle}; e.g.,][]{Brazzini25, Rusakov26, Matthee26}. The broadening has no information about kinematics (e.g., due to a virial broad-line region) and is produced entirely by radiative transfer.
    }
\label{fig:electronscattering}
\end{figure}

\begin{figure*}
    \centering
\includegraphics[width=0.9\linewidth]{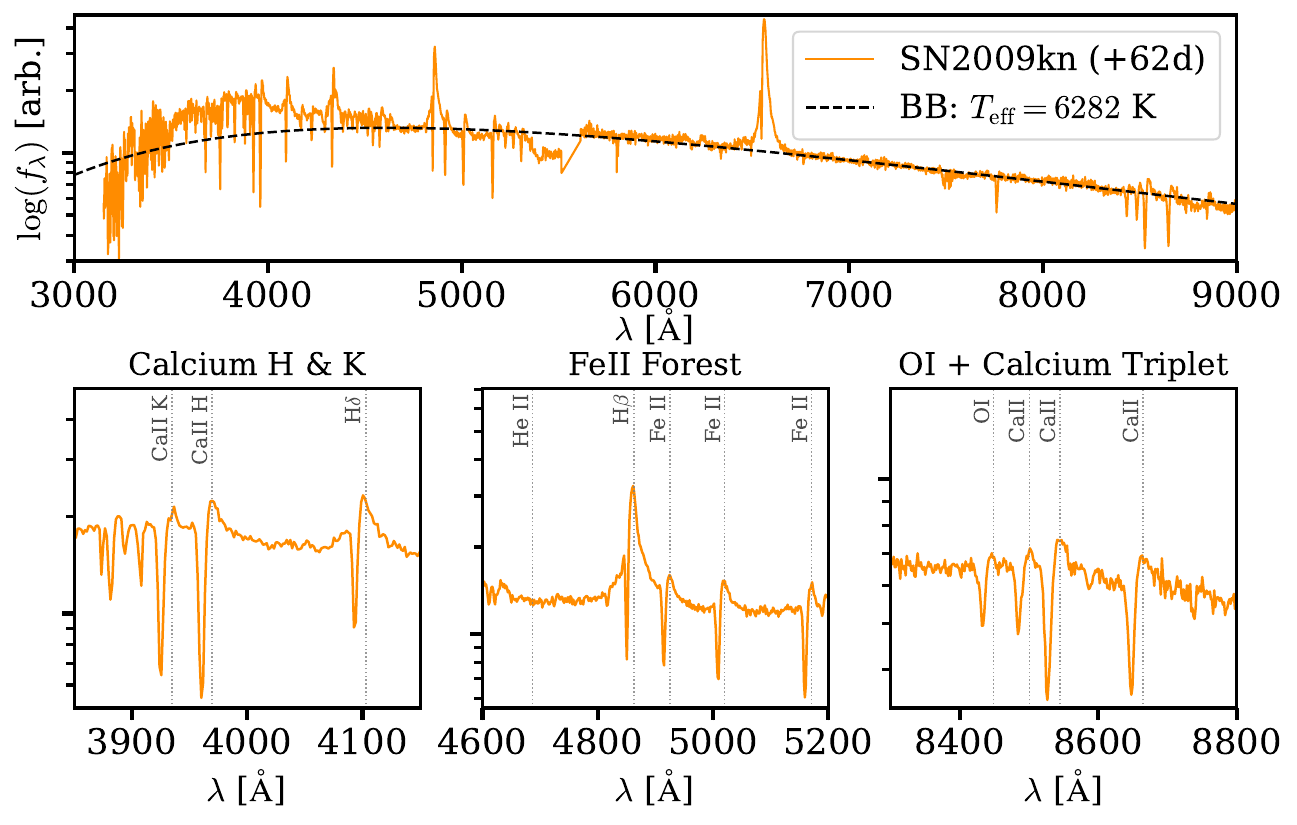}
    \caption{\textbf{LRD-like pseudo-photospheric features highlighted in SN2009kn (orange; $R\approx8800\ (5500)$ at $\gtrsim5500$\AA\ ($\lesssim5500$\AA); \citealt{Kankare12}).} \textbf{Top:} The overall continuum shape at $\gtrsim4500$\AA\ is well-described by a single-temperature blackbody akin to LRDs \citep[e.g.,][]{degraaff25pop, Sun26}. LRDs have sharper Balmer breaks than observed in SNe, but it is interesting to note that a Balmer break and a rollover in the continuum is apparent. \textbf{Bottom:} Quintessential ``stellar" absorption features observed in LRDs such as Ca H \& K absorption (left; e.g., \citealt{deugenio25irony}) and the Calcium triplet (right; e.g., \citealt{Lin25}) are highlighted in these panels. The \ion{Fe}{2} forest (center; e.g., \citealt{Torralba25IFU}) and \ion{O}{1} (right; e.g., \citealt{Tripodi25}) are relatively uncommon among galaxies/AGN but are ubiquitous among Type IIn SNe and LRDs. Note that the P-Cygni absorption is more apparent than in typical LRDs -- there is no host galaxy infilling, and the velocity offset of absorption from line-center is larger (see e.g., Fig. \ref{fig:lineprofiles} in H$\alpha$).
    }
\label{fig:fullspec}
\end{figure*}

\begin{figure*}
    \centering
\includegraphics[width=\linewidth]{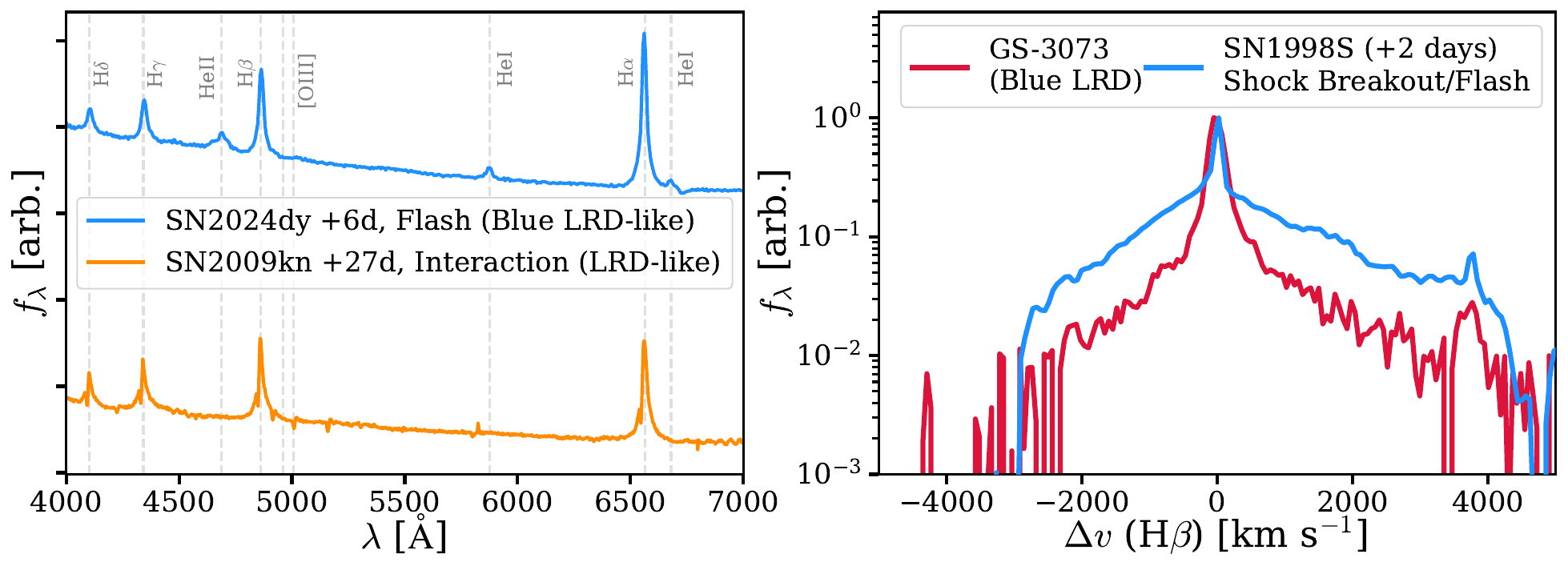}
    \caption{\textbf{Phases of Type IIn SNe evolution may help interpret sub-classes of LRDs.} \textbf{Left:} The initial shock breakout or ``flash" phase of IIn SNe (blue; \citealt{Ihanec24}) is marked by high ionization lines such as \ion{He}{2} with the Balmer lines showing no absorption. These are the defining features of the bluest LRDs \citep[e.g.,][]{Matthee26, Brazzini26}. In the much longer-lived interaction phase (orange; \citealt{Kankare12}), the Balmer lines have P-Cygni profiles and high ionization lines are absent, reminiscent of typical LRDs. \textbf{Right:} An H$\beta$ spectrum of the flash phase \citep[][]{Shivvers15} is contrasted against an archetypal blue LRD, GS-3073 \citep[e.g.,][]{Ubler23, Brazzini26, Matthee26}. The Keck/HIRES spectrum is degraded to the JWST/NIRSpec resolution for the comparison. The lines both have extremely narrow cores with broad electron scattering wings -- broader in the case of the SNe whose ``cold wind" is more ionized. There is no sign of Balmer absorption (contrast against bottom-center panel of Fig. \ref{fig:fullspec} above). 
    }
\label{fig:snphases}
\end{figure*}

Here we highlight a constellation of defining empirical features that occur across LRDs, the GE, and IIn SNe. Unless stated otherwise, comparisons to the GE and IIn SNe in this section correspond to the interaction-dominated phase of these phenomena where these objects spend a significant amount of time -- few months to years for IIn SNe \citep[e.g.,][]{Taadia13, Hiramatsu24, Ransome25}, and two decades for the GE \citep[e.g.,][]{Smith18}. 

While each of the following features, individually, may be found in some or the other class of AGN or galaxy, it is remarkable that all of them occur \textit{simultaneously} in these objects. That these unusual features go hand-in-hand (e.g., a V-shaped SED in a point-source virtually guarantees broad-lines; \citealt{Hviding25}) points to a unified underlying physical mechanism. The vast zoo of AGN may show a handful of LRD-like properties, but never display the entire constellation -- e.g., Balmer absorption is seen in a handful of AGN \citep[e.g.,][]{Shangguan26}, but they are variable \citep[e.g.,][]{Park26}. In what follows, we list and discuss the empirical similarities between LRDs and enshrouded eruptions:

\begin{itemize}
    \item \textbf{Blackbody-like Optical Continua with $T_{\rm{eff}}\approx4000-7000$ K:} The GE \citep[e.g.,][]{Rest12, Prieto14, Smith18}, LRDs \citep[e.g.,][]{degraaff25pop, Umeda25, Sun26}, and Type IIn SNe \citep[e.g.,][]{Taadia13, Hiramatsu24} show blackbody-like continua peaking in the rest-optical or NIR (Figure \ref{fig:hrd}). The pseudo-photosphere in these objects is the surface we are seeing in these blackbody-like continua -- the actual surfaces of the central engines (or in SNe, radioactive material from explosions) can be orders of magnitude hotter, but they are completely buried within the opaque wind obscuring them. The characteristic $T_{\rm{eff}}$ of these ``recombination photospheres" \citep[e.g.,][]{Owocki16} is an outcome of hydrogen opacity physics -- this is the $T_{\rm{eff}}$ range where recombination produces a steep opacity gradient, as has also been noticed in the LRD literature \citep[e.g.,][]{Kido25, Liu25BB, degraaff25pop}. Further validating this interpretation, (pseudo-)photospheric absorption features such as the Calcium Triplet (CaT), Calcium H\&K, and Sodium D  are observed in LRDs \citep[e.g.,][]{Lin25, deugenio25irony, Liu26TLUSTY} as well as IIn SNe (see Fig. \ref{fig:fullspec}; e.g., \citealt{Kankare12}). These are quintessential features of stellar spectra -- e.g., the \textit{Gaia} RVS spectrograph \citep[][]{GaiaDR3} was designed to measure CaT for millions of stars. Note that the detection of these absorption features shows that a pseudo-photospheric continuum may co-exist with the nebular continuum seen in the form of e.g., Paschen jumps in models and observations of LRDs \citep[e.g.,][]{degraaff25, Sneppen26Paschen, Sun26}.
    
    \item \textbf{Molecules tracing cold gas:} The GE and IIn SNe show molecular features such as CO overtones \citep[e.g.,][]{Fassia01} and CN bands \citep[e.g.,][]{Prieto14} that imply the presence of cold ($<4500$ K) gas. Water in LRDs \citep[e.g.,][]{Wang26} has been similarly interpreted to be tracing $<3000$ K material. At face value, such cold gas is at odds with a $T_{\rm{eff}}\approx5000$ K photosphere \citep[][]{Prieto14, Wang26}. However, here it is important to note a key difference between stellar and pseudo-photospheres -- a large temperature and density gradient exists above the pseudo-photosphere. As summarized by \citet[][]{Davidson16} in the context of the GE: \textit{``Casual analogies such as `reminiscent of an F-type supergiant' can be justified, but formally one should not assign a stellar spectral type to the absorption features of an opaque outflow. Imagine, for instance, a star with $T_{\rm{eff}} = 6500$ K, compared to an opaque wind with the same photosphere temperature. The star’s atmosphere has practically no material with $T<5200$ K, but outer parts of the wind can have $T<4500$ K. Hence, the wind can form absorption lines much cooler than anything in the star’s spectrum."} 
    
    \item \textbf{What you see is what you get -- $L_{\rm{opt}}$ comparable to $L_{\rm{bol}}$:} A hallmark of LRDs is their X-ray and IR weakness. The bolometric luminosity is emerging almost entirely in the rest-optical and NIR such that $L_{\rm{bol}}/L_{\rm{opt}}\approx3$ \citep[e.g.,][]{Yue24, Greene25, Casey25}. This is indeed the case for the GE and Type IIn SNe -- see blackbody-based $L_{\rm{bol}}$ estimates in \citet{Rest12, Hiramatsu24} similar to LRD $L_{\rm{bol}}$ estimates pioneered by \citealt{degraaff25pop} and validated in \citet[][]{Greene25}. The dense envelopes of these objects ($\gtrsim10^{24-25} \ \rm{cm^{-2}}$; e.g., \citealt{Dessart16}) are mostly opaque to X-rays created by their shocks, which are reprocessed into optical and IR light \citep[e.g.,][]{Chevalier12, Smith13} -- indeed, $<5\%$ of Type IIn SNe surveyed for X-rays have shown detections \citep[e.g.,][]{Chandra18}. Dust production (and hence non-negligible FIR flux) is a feature of Type IIn SNe, but this is in later stages and does not account for the red optical colors of the objects that are set by gas physics \citep[e.g.,][]{Fox11,Medler25}.

    \item \textbf{Hydrogen line profiles shaped by electron scattering and neutral gas absorption:} LRDs are marked by strong H$\alpha$ emission with exponential wings \citep[e.g.,][]{Torralba25IFU, Rusakov26, Sneppen26} accompanied by Balmer absorption \citep[e.g.,][]{Matthee24, deugenio25twice, Davis26}. The line cores often resemble massive star winds with e.g., P-Cygni profiles \citep[e.g.,][]{Matthee26}. Balmer lines with similar traits have been observed in the GE \citep[][]{Smith18} and are a defining signature of Type IIn SNe \citep[e.g.,][]{Chugai04, Kiewe12, Taadia13}. Furthermore, subtle features of LRD line profiles are seen in IIn SNe (see Fig. \ref{fig:hasubtle}). Balmer absorption \textit{increases} down the Balmer series like a stellar atmosphere \citep[e.g.,][]{Taadia13, Naidu25BHstar}, contrary to expectations from oscillator strengths. The velocity offset of absorption \textit{decreases} down the series -- e.g., the H$\beta$ absorption is generally less blueshifted than the H$\alpha$ absorption \citep[e.g.,][]{Kankare12, Matthee26}. The electron scattering wings get broader as one goes from H$\alpha$ to H$\beta$ to H$\gamma$ (see Fig. \ref{fig:electronscattering}; e.g., \citealt[][]{Dessart09, Brazzini25}). And finally, an ``intermediate" narrow component from the central engine, distinct from the host galaxy, is detected \citep[e.g.,][]{Smith17, Matthee26}. These are signatures of lines forming in a medium that is dense enough to produce Balmer absorption, neutral enough to produce a P-Cygni line core, and ionized enough to produce broad electron scattering wings. Evidence for a bulk outflow comes from the P-Cygni profiles and the velocity offsets between Balmer lines (higher order lines are absorbed closer to systemic). The higher order lines form at deeper depths -- as a result, the lines experience different column depths of electrons and dense gas, therefore showing different velocity offsets and line profiles \citep[e.g.,][]{Dessart09, Sneppen26}. We emphasize how remarkable it is that \textit{all} these Balmer line trends are seen in some form in IIn SNe and the GE -- consider that just Balmer absorption by itself is so rare that there are only a handful of examples among all known galaxies and AGN barring LRDs \citep[e.g.,][]{Hall07, Hall13, Williams17, Park26, Shangguan26}. 
    
    \item\textbf{Extreme Balmer Decrements from Dense Gas, not Dust:} While Balmer decrements for the GE are yet to be reported, it is not uncommon for Type IIn SNe to show the highly unusual H$\alpha$/H$\beta>10$ \citep[e.g.,][]{Aretxaga99, Kankare12, Fransson14} routinely observed in LRDs \citep[e.g.,][]{Brooks25Balmer, degraaff25pop,Nikopoulos25,Sun26, Lin26}. The steep Balmer decrement arises from radiative transfer in dense gas \citep[e.g.,][]{Xu92} similar to mechanisms proposed for LRDs \citep[e.g.,][]{Chang25, Yan25, Torralba25IFU}. In particular, in dense, optically thick line forming conditions collisional excitation and resonant scattering cascades boost H$\alpha$, while repeated scattering and line trapping diminish H$\beta$. Note that this mechanism to produce steep Balmer decrements arises from the same medium responsible for the Balmer break as well as the Balmer absorption, instead of invoking an additional component such as dust \citep[e.g.,][]{Madau26, Martins26, Perez-Gonzalez26}. Also note that the required dust law to reconcile the Balmer series would be unprecedented \citep[e.g.,][]{Nikopoulos25}.

     \begin{figure*}
    \centering
\includegraphics[width=0.8\linewidth]{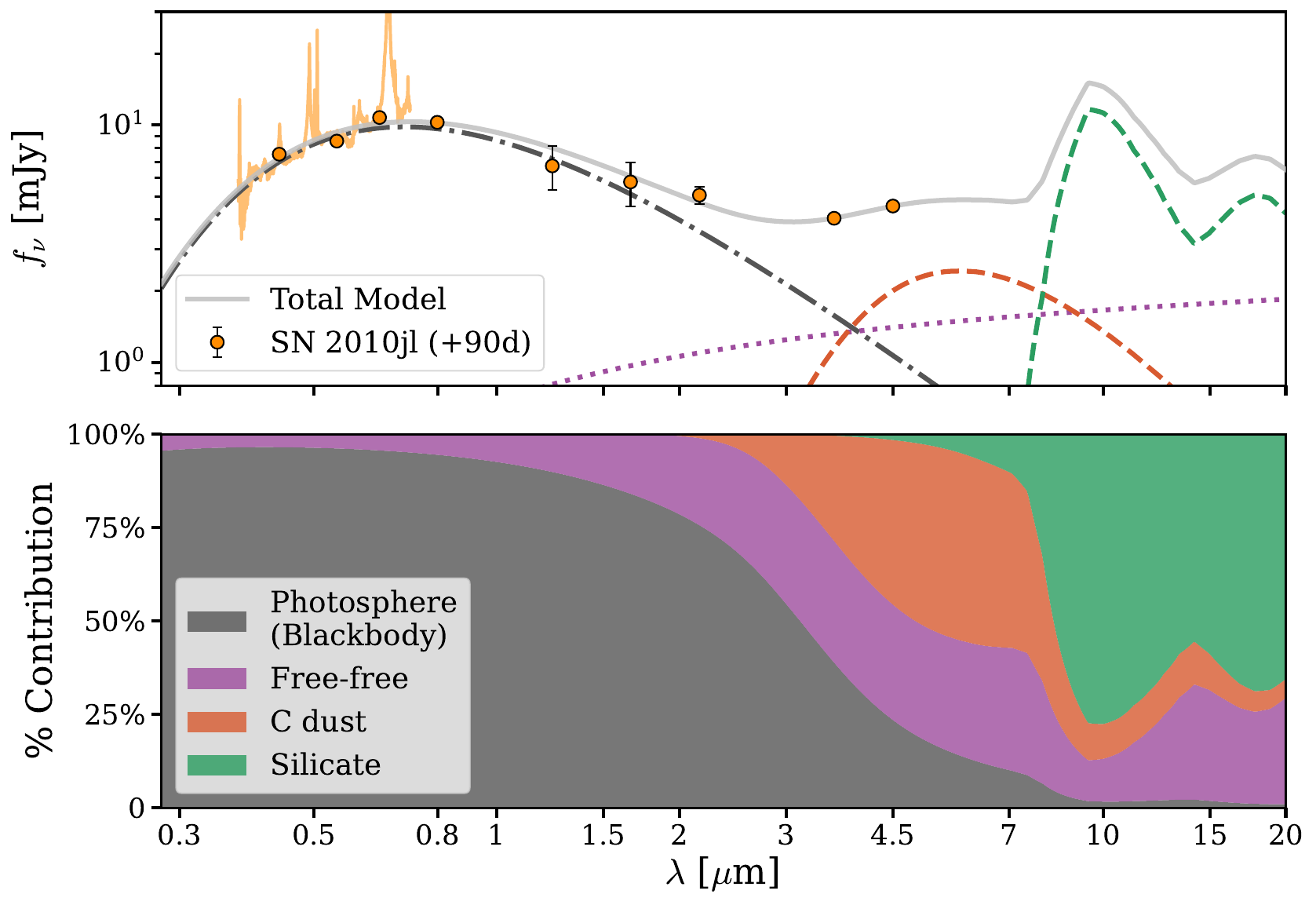}
    \caption{\textbf{The MIR spectrum of Type IIn SNe may help illuminate the MIR of LRDs.} We show an illustrative fit to SN2010jl \citep[][]{Andrews11} with the typical components included in such modeling \citep[e.g.,][]{Derkacy26, Medler25}. The key features in the MIR include a blackbody representing the pseudo-photosphere ($T_{\rm{eff}}=7300$ K; gray), free-free emission ($T=20,000$ K; purple), Carbon dust ($T=550$ K; dark orange), and Silicate dust ($T=350$ K; green).
    }
\label{fig:freefree}
\end{figure*}

    \item \textbf{Collisional Excitation, Fluorescence, and Pumping seen in \ion{Fe}{2} forest and \ion{O}{1}:} A distinctive feature of LRDs is a rich forest of \ion{Fe}{2} lines spanning the UV through the IR and strong \ion{O}{1} 8446, 11290 \AA\ emission \citep[e.g.,][]{Kokorev25glimpsed, deugenio25irony, Torralba25IFU, Lin25, degraaff25pop, Tripodi25}. These features arise from optically thick gas with  $T_{\rm{e}}\approx6000-7000$K that is efficient at Lyman-$\alpha$/Lyman-$\beta$ pumping and collisional excitation \citep[e.g.,][]{Grandi80, Sigut98, Sigut03, Torralba25IFU, Dias26, Martins26}. Strong \ion{Fe}{2} transitions and \ion{O}{1} are observed in the GE \citep[][]{Smith18} and routinely in IIn SNe \citep[e.g.,][]{Pastorello02, Dessart09}.
    
    \item \textbf{Prominent, P-Cygni \ion{He}{1}:} The 5877~\AA, 7067~\AA\,and 10830~\AA\ \ion{He}{1} lines are among the strongest in LRDs \citep[e.g.,][]{Matthee24, Wang24, Kokorev25glimpsed}, and are seen in the GE \citep[][]{Smith18} and Type IIn SNe \citep[e.g.,][]{Kiewe12}, often with dramatic, textbook P-Cygni line profiles \citep[e.g.,][]{Fassia01}. Intriguingly, the metastable, triplet 10830~\AA\ line in LRDs, almost without exception, shows a similarly deep P-Cygni profile reminiscent of massive star winds \citep[e.g.,][]{NM24,Juodzbalis24, Loiacono25, Wang24, Kokorev25glimpsed}. The kinematics of the 10830~\AA\ line in LRDs are relatively uncontaminated by host galaxy infilling -- the central engine dominates the light at these wavelengths, and \ion{He}{1} in the hosts is relatively weak \citep[e.g.,][]{Sun26}. On the other hand, interpreting H$\alpha$ absorption is extremely challenging and sensitive to modeling choices and spectral resolution of the observations \citep[e.g.,][]{Matthee26, Davis26}. Therefore, it is heartening that the outflowing absorption seen in the Balmer lines is supported by \ion{He}{1}.

    \item \textbf{UV -- Ly$\alpha$, high ionization lines, UV continuum:} While in the typical LRD, the majority of UV light arises from the host galaxy, there are significant contributions from the central engine as well, ranging from $\approx0\%$ in the faintest LRDs to $>50\%$ in the brightest ones \citep[e.g.,][]{Sun26, Asada26, Ando26}. Further, there are a handful of LRDs in which tentative detections of high-ionization UV lines and broad Ly$\alpha$ have been reported \citep[e.g.,][]{Tang25, Tang26, Ji26} -- though note, in the typical LRD Ly$\alpha$ emission is consistent with arising entirely from the host \citep[e.g.,][]{Ji26LyA}. While UV observations of IIn SNe are rare, these features have been observed in a small number of sources -- e.g., SN2010jl \citep[][]{Fransson14}. The interpretation is that in these cases photons from the forward shock (e.g., during the early ``flash" or shock breakout) are able to escape into and photoionize the cold wind \citep[e.g.,][]{Chandra18}. Furthermore, LBV winds as in the GE have been modeled to be clumpy and porous \citep[e.g.,][]{Shaviv00,Davies05, Weis20} -- supported by e.g., the discovery of the \citet[][]{Weigelt86} ``blobs" around $\eta$ Car -- possibly facilitating Ly$\alpha$ escape through relatively lower density sightlines in the cold wind \citep[e.g.,][]{Tang26, Ji26}.

    \item \textbf{A dearth of [\ion{O}{3}]:} Several lines of evidence suggest the [\ion{O}{3}] doublet in the rest-optical ($\lambda\lambda4960, 5008$\AA), often the strongest line in high-redshift galaxy spectra \citep[e.g.,][]{Matthee23, Endsley24bursty, Meyer24}, is conspicuously weak among the central engines of LRDs. This is particularly striking in the LRDs where the central engine accounts for almost all the rest-optical light \citep[e.g.,][]{Naidu25BHstar, degraaff25, Torralba26panbhstar}. Further, empirical models that assume the majority of [\ion{O}{3}] is produced by the host galaxy are able to reproduce most of the diversity of LRDs \citep[e.g.,][]{degraaff25pop,Barro25,Sun26}. In our analogy, the weakness of [\ion{O}{3}] is a generic feature of Type IIn SNe (Figs. \ref{fig:fullspec}, \ref{fig:snphases}). This is an outcome of the gas densities involved, which exceed the critical density for the [\ion{O}{3}] doublet by a few orders of magnitude ($n_{\rm{H}}\approx10^{6} \rm{cm^{-3}}$ vs. 10$^{9-12}$; e.g., \citealt{Dessart16}). Further, the ionization field for the bulk of the material is low across the vast span of IIn spectral forming regions \citep[e.g.,][]{Dessart16}. In the GE spectra, unfortunately, [\ion{O}{3}] is contaminated by photoionization from the ISM structures that the light echoes are reflected from \citep[e.g.,][]{Smith18}.

    \newcommand{\pbF}[1]{\parbox[t]{0.30\linewidth}{\raggedright\rule{0pt}{2.6ex}#1\vspace{1.4mm}}}
\newcommand{\pbI}[1]{\parbox[t]{0.65\linewidth}{\raggedright\rule{0pt}{2.6ex}#1\vspace{1.4mm}}}
\begin{deluxetable*}{|c|c|}
\tabletypesize{\footnotesize}
\tablecaption{Summary of empirical features shared between LRDs and Enshrouded Eruptions\label{tab:bhstar_features}}
\tablehead{
\colhead{\pbF{\centering\textbf{Feature}}} & \colhead{\pbI{\centering\textbf{Interpretation}}}}
\startdata
\hline
\pbF{$T_{\rm{eff}}\approx5000$ K Blackbody-like Continuum} & \pbI{Hydrogen opacity physics regulating a recombination/wind/pseudo photosphere.} \\ \hline
\pbF{Molecular features like H$_{\rm{2}}$O, CO, CN} & \pbI{Imprinted by cold ($T<3000-4000$ K) layers in the outskirts of the wind.} \\ \hline
\pbF{Optical luminosity comparable to bolometric luminosity} & \pbI{X-rays, FUV absorbed and reprocessed into optical and IR light; sub-dominant but non-zero contribution from MIR/FIR dust and free-free emission.} \\ \hline
\pbF{P-Cygni line profiles (e.g., in \ion{He}{1}, H$\alpha$, H$\beta$)} & \pbI{Formed in the cold wind, which is an optically thick, partially ionized+neutral, net outflowing medium.} \\ \hline
\pbF{Broad exponential wings (across all lines to varying extent)} & \pbI{Electron scattering in partially ionized cold wind, with extent of scattering varying with line formation depth.} \\ \hline
\pbF{Extreme Balmer decrement (typical H$\alpha$/H$\beta>10$)} & \pbI{Collisional excitation and Balmer scattering which converts H$\beta$ into Pa$\alpha$ and H$\alpha$.} \\ 
\hline
\pbF{Strong \ion{O}{1} emission} & \pbI{Lyman-$\beta$ pumping, efficient when Balmer lines are optically thick.} \\ 
\hline
\pbF{\ion{Fe}{2} transitions (UV, optical, IR)} & \pbI{Lyman pumping, soft ionizing spectrum, $T\approx6000-7500$ K gas consistent with conditions implied by the pseudo-photosphere.} \\ 
\hline
\pbF{Dearth of [\ion{O}{3}] 5008\AA} & \pbI{Density of the wind far higher than the critical density ($>10^{9}$ cm$^{-3}$ versus $\approx10^{6}$ cm$^{-3}$), low ionization field across the CSM.} \\ 
\hline
\pbF{Rare tail of blue, high-ionization LRDs} & \pbI{Analogous to short-lived phases of SNe evolution (e.g., shock breakout) where a larger fraction of hard ionizing photons traverse the cold wind.} \\ 
\hline
\pbF{MIR Excesses} & \pbI{Free-free emission from the wind and dust emission from \textit{newly forming} dust in the $T<3000$ K layers of the wind.}\\
\hline
\pbF{UV: Ly$\alpha$, high ionization lines (e.g., \ion{N}{5})} & \pbI{Flash photoionization and/or radiative transfer through a clumpy medium with a mixture of low and high column density regions.}
\enddata
\end{deluxetable*}

\item \textbf{Rare blue LRDs as stages of SNe evolution?} A small minority of broad-line emitters at $z\gtrsim5$ resemble LRDs in all respects (e.g., X-ray faint, not variable, exponential lines, strong \ion{He}{1}), except they lack Balmer breaks, Balmer absorption, and show high ionization lines, specifically \ion{He}{2} \citep[e.g.,][]{Ubler23,Brazzini26, Matthee26}. Note that these few LRDs must not be conflated with those where light from the central engine is simply diluted by an extended blue host galaxy \citep[e.g.,][]{Sun26, Rinaldi26, Billand26} -- for instance, \citet[][]{Sun26} demonstrate V-shaped SED selections are largely sensitive to sources where the central engine contributes $\gtrsim60\%$ of the optical light. The \ion{He}{2} LRDs may be powered by the same central engines as other LRDs but are perhaps observed at different viewing angles and/or at different evolutionary stages such that the gas density along the line of sight is lower \citep[e.g.,][]{Matthee26, Madau26, Sneppen26lbd}. Intriguingly, some relatively transient stages of IIn SNe evolution may be mapped to such sources (Fig. \ref{fig:snphases}). In particular, high ionization lines like \ion{He}{2} and no Balmer absorption (but exponential wings) are observed during the early flash/shock breakout before the recombination photosphere forms and/or during late stages with lower \ion{H}{1} column densities when the circumstellar medium has been thinned out \citep[e.g.,][]{Shivvers15}.

    \item \textbf{MIR -- Dust + Free-Free Emission:} In some LRDs, particularly those at $z\approx0$, clear excesses above the photospheric continuum are detected in the MIR \citep[e.g.,][]{Lin25, Lin26, Ji25lol, Park26}. This is typically attributed solely to dust \citep[e.g.,][]{delvecchio25, Brazzini26, Perez-Gonzalez26}.  Such excesses are observed in Type IIn SNe, and are modeled as a combination of free-free emission as well as dust emission (Fig. \ref{fig:freefree}). Free-free emission may be particularly important in LRDs -- based on fundamental wind physics, the strength is a function of the rate at which mass is being added to the wind ($\dot{M}$; ``mass-loss rate" in a stellar wind context) and the asymptotic, terminal wind velocity \citep[$v_{\rm{\infty}}$, e.g.,][]{Wright75}. The $\dot{M}$ for the central engines of LRDs may be substantial, as suggested e.g., by models where their envelopes form from pulsations of supermassive stars \citep[e.g.,][]{Nandal26}. And so like Type IIn SNe, LRDs may be expected to be free-free emitters as well \citep[e.g.,][]{Derkacy26, Dessart26}. The infrared spectra also show contributions from \textit{newly forming} dust \citep[e.g.,][]{Shahbandeh25}. Analogous to IIn SNe, LRDs may be prodigious dust producers as well (see \S\ref{sec:dust}). It is critical to note that the red optical spectra are not shaped by this dust, but by gas -- e.g., the steep Balmer breaks (typical strength of $\approx7$, $\approx2\times$ the stellar population maximum; \citealt{Sun26}) as well as Balmer decrements \citep[e.g.,][]{Nikopoulos25} cannot be explained by dust.
\end{itemize}

We end this section by noting a few of the most significant differences between LRDs and these stellar phenomena. As we will discuss in detail in the following sections, most of these differences may be explained by ``scaling up" to match the LRDs' larger mass/energy/radial scale or appealing to distinct properties of LRD central engines (e.g., accreting black holes). Most basically, from duty cycle arguments it appears galaxies spend $\approx10-20$ Myrs in the LRD phase \citep[e.g.,][]{Sun26} -- $\approx10^{6}\times$ higher than the GE's two decades. There are also differences with respect to LBV and SNe spectra. Balmer breaks are often observed among IIn SNe (see Fig. \ref{fig:fullspec}), but they are not quite as steep as in LRDs \citep[e.g.,][]{Naidu25BHstar, degraaff25}. Balmer absorption in LRDs extends to redshifted velocities \citep[e.g.,][]{Torralba26panbhstar, Davis26, Lin26} -- this is rarely seen among enshrouded eruptions, but plausible explanations include pulsation \citep[e.g.,][]{Wolf90, Zhang25cepheid}, rotation \citep[e.g.,][]{Torralba26panbhstar}, or external accretion \citep[e.g.,][]{Kido25}.

    \begin{figure*}
    \centering
\includegraphics[width=0.9\linewidth]{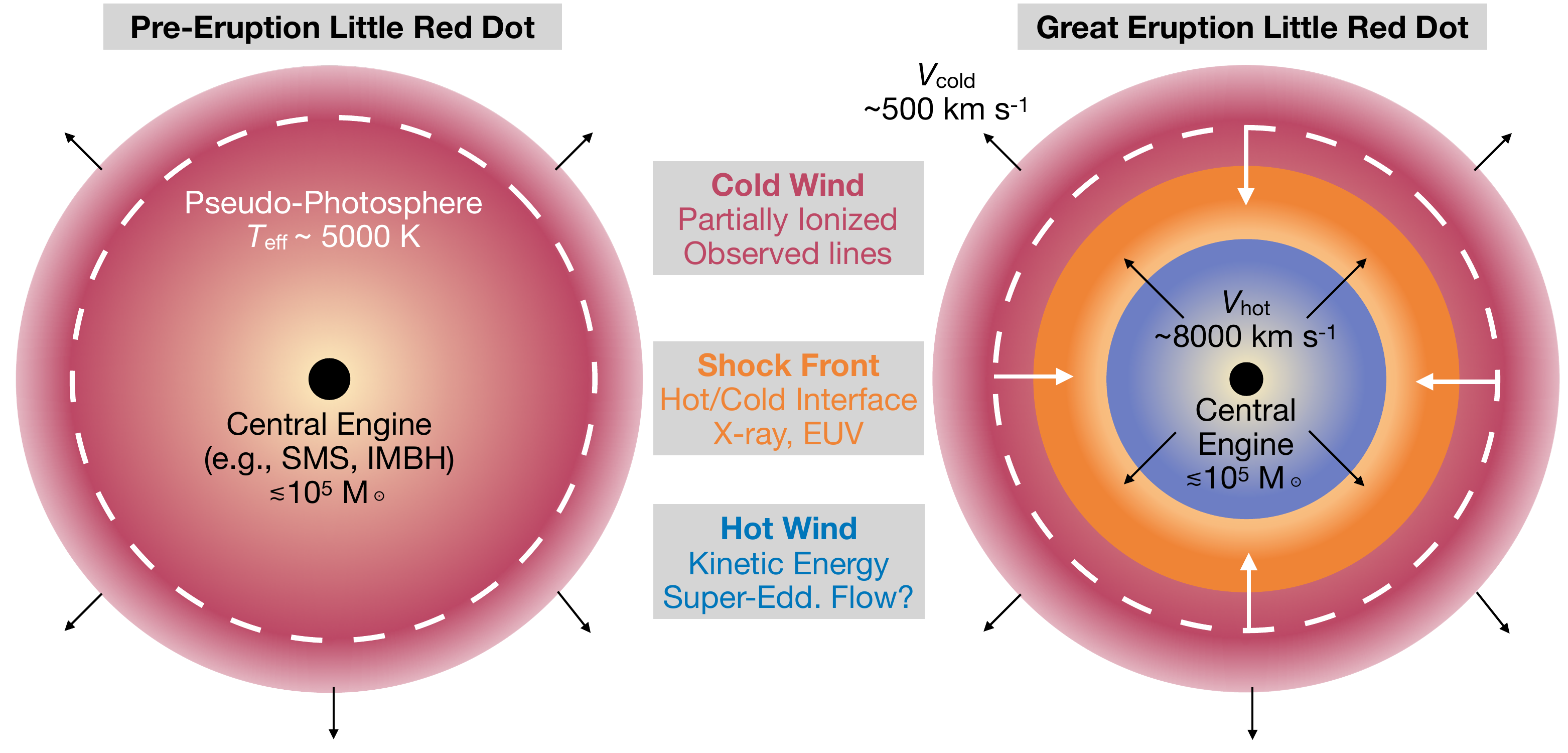}
    \caption{\textbf{Schematic of the proposed LRD scenario inspired by Type IIn SNe and the GE}. Figure modeled after scenarios in \citet[][]{Smith13, Smith17, Smith18}. \textbf{Left:} Prior to the LRD phase, the central engine drives a slowly outflowing envelope of gas that we refer to as the ``cold wind" (crimson with outward arrows). An optically thick pseudo-photosphere (white dashed lines) obscures the hot surface of the star and gives it the appearance of a $\approx5000$ K surface. Such a ``mini-eruption" phase is observed prior to the main explosion in Type IIn SNe/GE-like eruptions \citep[e.g.,][]{Ofek14}. \textbf{Right:} The central engine drives hot winds (blue) that crash into the surrounding cold wind. Shocks are produced in the interface between hot and cold winds (orange), and the kinetic energy is converted highly efficiently into radiation by this wind interaction mechanism. The cold wind on top of the shock front is (partially) ionized by the forward shock. The pseudo-photosphere recedes into the interface between the winds, and then into the hot wind over time \citep[e.g.,][]{Smith17}. Emission lines, the photospheric continuum, nebular emission, Balmer absorption arise from the partially ionized layers in the cold wind -- the cold wind is effectively what has been modeled in e.g., \texttt{CLOUDY} models for LRDs \citep[e.g.,][]{Naidu25BHstar, Torralba25IFU, Yan25} and other models that share a similar ionization structure \citep[e.g.,][]{Sneppen26}.}
\label{fig:schematic}
\end{figure*}

\section{The Physics of Little Red Dots as The Physics of Enshrouded Eruptions}
\label{sec:physics}

\subsection{Proposed Scenario: LRDs as ``Recurring Eruptions"}

Given their striking similarities, in this section we explore a scenario where LRDs are powered by similar-spirited physics as enshrouded eruptions \citep[e.g.,][]{Dessart09, Smith13,  Smith18}. Our approach is to stick as closely as possible to the stellar exemplars and adapt them to LRD observables (Fig. \ref{fig:schematic}). We do not claim this to be \textit{the} unique physical scenario that explains LRDs, but \textit{a} scenario that deserves consideration given the constellation of non-trivial empirical parallels, which to our knowledge are not observed in any other class of object (\S\ref{sec:empirical}). We also emphasize that we are not claiming LRDs to be \textit{literal} IIn supernovae or great eruptions, but that analogous physics may govern them (e.g., radiation from a central engine reprocessed by dense enshrouding gas). Indeed, crucial differences exist: the central engine is likely orders of magnitude more massive, and is likely to be persistent, e.g., aided by accretion onto an SMS/BH or a continuum driven wind and not just a one-time explosion or eruption. Keeping the analogies to massive star phenomena in mind, a possible scenario for LRDs is as follows:

\begin{itemize}
    
    \item \textbf{The Pre-LRD Phase -- The Central Engine is Enshrouded by a Dense Wind:} The decades preceding the GE \citep[e.g.,][]{Smith18} and archival observations of Type IIn SNe progenitors \citep[e.g.,][]{Ofek14} provide evidence for ``mini-eruptions" whose hydrogen-rich ejecta account for the dense CSM enveloping the central engine. Possible mechanisms for creating such a shroud of material around LRD central engines include e.g., super-Eddington flows or disk winds from accreting black holes \citep[e.g.,][]{Liu25BB, Lam26, Torralba26panbhstar}, pulsational mass loss or eruptions from supermassive stars \citep[e.g.,][]{Nandal26}.
    
    \item \textbf{The Enshrouding Envelope may be Kinematically Complex:} LRDs show a diversity of absorption signatures that are not always pure P-Cygni outflows -- e.g., the absorption in H$\alpha$ may extend to red velocities \citep[e.g.,][]{Davis26}, though note inferring this robustly is challenging due to infilling from the host galaxy \citep[e.g.,][]{Matthee26}. Such complex signatures may be the result of turbulence \citep[e.g.,][]{Hopkins24}, a launching surface that is rotating as in the case of an accretion disk wind \citep[e.g.,][]{Torralba26panbhstar}, a rotating supermassive star \citep[e.g.,][]{Haemmerle21}, or a combination thereof (accretion disk around a rotating supermassive star; e.g., \citealt{Zwick26}) . 
    
    \item \textbf{Formation of a Hydrogen Recombination Pseudo-Photosphere in the Enshrouding Wind:} The hot surface of the central engine is fully obscured by the dense, optically thick envelope such that the $\tau=2/3$ surface (i.e., the ``pseudo-photosphere") lies \textit{within} the wind \citep[e.g.,][]{Nandal26, Chisholm26}. A consequence is that a blackbody-like continuum is observed with $T_{\rm{eff}}$ far lower than the temperature at the surface of the central engine. The $T_{\rm{eff}}$ has a propensity for $\approx4000-7000$~K (Fig. \ref{fig:hrd}) set by hydrogen opacity physics in a so-called ``recombination photosphere" \citep[e.g.,][]{Dessart09, Dessart11, Owocki16, Kido25, Liu25BB, Chen26}.
    
    \item \textbf{A Recurring ``Fast Wind" from an Enshrouded Central Engine:} While completely enshrouded, the central engine injects energy that we describe as a ``fast wind" \citep[e.g.,][]{Quataert16,Owocki17, Smith18}. In the case of the GE the precipitating event for the fast wind was likely a stellar merger \citep[e.g.,][]{Hirai21}, whereas in a Type IIn SN this is an explosion from the core collapse of a massive star. For LRDs, we speculate that the winds could arise from e.g., accretion onto a black hole which launches episodic outflows \citep[e.g.,][]{Liu25BB} or from a supermassive star that launches successively stronger winds \citep[e.g.,][]{Nandal26}. Critically, these outflows crash into the existing dense gas envelope. We envision this to be a ``recurring eruption" in that the central engine is a relatively steady and long-lived source that may keep launching winds versus a one-time terminal explosion like an SN. 
    
    \item \textbf{Radiative Shocks and Fast/Slow Wind Interaction Contribute to LRDs' Immense Luminosity:} As the radiative shock from the interaction between the fast and slow winds plows through the dense medium, kinetic energy is converted into radiation highly efficiently \citep[e.g.,][]{Smith17}. Indeed, CSM interaction is the key mechanism by which the GE maintained an $L_{\rm{bol}}/L_{\rm{edd}}\gtrsim5$ for over two decades, and is the reason why some of the brightest known transients are interacting transients \citep[e.g.,][]{Smith18}. As we will discuss in \S\ref{sec:bhmasses}, CSM interaction can provide such a substantial boost to the luminosity that even very low-mass central engines become viable. 
\end{itemize}

Here we have outlined what it would take to explain LRDs as objects powered by similar physics as the GE and IIn SNe. Of course, some key differences may exist. For example, if the central engines are black holes, accretion may contribute to the ionizing spectrum in lieu of or in addition to the wind interaction mechanism and the resulting reprocessed SED could still look very similar \citep[e.g.,][]{Naidu25BHstar,Ji25BT, Torralba25IFU}. As we will discuss in \S\ref{sec:bhmasses} and \S\ref{sec:discussion}, it is extremely challenging to peer past the pseudo-photosphere to directly probe the central engine.

\subsection{The Invariance of LRDs from Delayed Photon Diffusion in a Dense Enveloping Medium}
\label{sec:diffusion}

The typical LRD shows no discernible signs of variability on year to decade timescales -- neither in continuum flux, nor in line flux, nor in line profiles \citep[e.g.,][]{Kokubo25, Burke25, Stone25var, Tee25, Park26, Liu26TWINKLE}. Intriguingly, one of the key features of enshrouded eruptions is their remarkable longevity. The GE was bright for almost two decades with a relatively flat light curve \citep[e.g.,][]{Smith18}. There are massive star eruptions that have lasted almost half a century \citep[e.g.,][]{Humphreys06}. Similarly, IIn SNe are among the longest-lived SNe (typically $\approx100$ days above $50\%$ of peak luminosity; e.g., \citealt{Hiramatsu24}), with some events remaining bright for several years \citep[e.g.,][]{Turatto93,Stritzinger12,Tartaglia20}. 

The photon diffusion timescale through the dense CSM is significant in these phenomena \citep[e.g.,][]{Hillier12}. We may approximately estimate this for LRDs as --
\begin{align}
    t_{\rm{diff}} &= \tau \left(\frac{R_{\rm{phot}}} {c}\right).
\end{align}

Here $\tau = \kappa_{\rm{es}} \rho R_{\rm{phot}}$, where $\tau$ is the optical depth,  $\kappa_{\rm{es}}$ is the electron scattering opacity, and $\rho$ is the average density which we approximate by assuming a uniform sphere. To estimate $\rho$, we assume $\approx1\%$ of the central engine mass is in the wind \citep[e.g.,][]{Nandal26}, which translates to $\approx1000 M_{\rm{\odot}}$ (see \S\ref{sec:bhmasses}). Alternatively, models of super-Eddington flows find $\rho\approx5\times10^{-12}$ g cm$^{-3}$ produces the continuum shape required to match typical LRDs \citep[e.g.,][]{Liu25BB}. We set $R_{\rm{phot}}=1300$ au based on the host-subtracted LRD stack of \citet[][]{Sun26}. This yields a diffusion timescale of $\approx10-700$ years for the range of wind masses and densities discussed above. 

Further, there exists an empirical correlation between the mass in the CSM and the lifetimes of IIn SNe (see Fig. 11 of \citealt{Ransome25}). Naively scaling the IIn SNe relation ($\log(M_{\rm{CSM}}/M_{\rm{\odot}})\approx 1.3 \log(t/\rm{days}) -2$) to LRDs with an envelope mass of $\approx1000 M_{\rm{\odot}}$ we find a duration of $\approx20$ years for the LRD phase. We are unaware of a similar scaling relation for GE-like events, but we note that the GE itself, with a CSM mass of $\approx10 M_{\rm{\odot}}$ \citep[e.g.,][]{Smith18} lasted $\approx20$ years, implying centuries or millennia for a system with $\approx100\times$ the CSM mass if a relationship similar to IIn holds here.

These numbers suggest that any change interior to the pseudo-photosphere will be smeared out over decades or even longer, explaining the LRDs' invariance. Furthermore, the central engines may be intrinsically invariant to begin with, as suggested by super-Eddington accretion models \citep[e.g.,][]{Secunda26, Liu26TWINKLE}.

\subsection{Energetics and Mass-loss Rates}
\label{sec:windparams}

If LRDs are powered by CSM interaction, it is useful to contextualize them among such phenomena. Here we present rough estimates for parameters typically computed for such objects. Following e.g., \citet[][]{Smith17}, the luminosity powered by the interaction between the hot and cold winds is often estimated as:

\begin{align}
    L_{\rm{bol}} &= \frac{1}{2} \dot{M} v_{\rm{hot}}^{2} \left(\frac{v_{\rm{hot}}}{v_{\rm{cold}}}\right). 
\end{align}

Here $\dot{M}$ is the mass loss rate from the central engine into the winds. The mass loss rate can be approximated by assuming spherical symmetry, that the wind is ionized, and that electron scattering is the dominant opacity source \citep[e.g.,][]{Piro20}:

\begin{align}
    \dot{M} &= \frac{4\pi v_{\rm{cold}} R_{\rm{phot}}\tau}{\kappa_{\rm{es}}}.
\end{align}

For the typical LRD, $L_{\rm{bol}}\approx10^{44}$ erg s$^{-1}$, $R_{\rm{phot}}=1300$ au \citep[][]{Sun26}, and the terminal velocity of the cold wind is $\approx500$ km s$^{-1}$ (see \S\ref{sec:vesc}). By definition, $\tau=2/3$ at $R_{\rm{phot}}$. This yields a mass loss rate of $\approx 0.3 M_{\rm{\odot}}\rm{yr}^{-1}$ and $v_{\rm{hot}}\approx8000$ km s$^{-1}$. The mass loss rate is an order of magnitude larger than seen in IIn SNe progenitors (consistent with the larger CSM mass in LRDs; see \S\ref{sec:diffusion}), while the velocity of the hot wind is comparable -- e.g., $\approx2500-10,000$ km s$^{-1}$ in typical IIn SNe and the GE \citep[e.g.,][]{Smith17, Smith18}.

Note that the estimates here are approximate order of magnitude calculations. In detail, eruptive mass loss and outflows from the central engine may be highly episodic and the wind structure (e.g., ionization, density, velocity, opacity profile) is likely far more complex \citep[e.g.,][]{Dessart16, Torralba26panbhstar, Martins26}. Also note that while duty cycle lifetimes indicate the host galaxies of LRDs must be observed in that state for $\approx10$ Myrs \citep[e.g.,][]{Sun26}, this is likely distributed across several LRD events (see \S\ref{sec:sms} for an estimate of $\approx30$) -- accounting for this makes the integrated mass lost more consistent with the densities discussed in \S\ref{sec:diffusion}. 

\subsection{The Co-Existence of Emission Lines and Cold Photospheric Continuum}
\label{sec:coexist}

LRDs simultaneously display a blackbody-like optical continuum alongside strong emission lines \citep[e.g.,][]{degraaff25pop, Sun26, Torralba26panbhstar}. The interpretation of the continuum as arising from a $\approx5000$ K (pseudo-)photosphere \citep[e.g.,][]{ Kido25, Santarelli25,Begelman25, Liu26TLUSTY, Roman-Garza26} has been challenged. If there is a cold photosphere ($T_{\rm{eff}}\approx5000$ K), then how are the lines being produced as this temperature is much too cold to produce a significant quantity of ionizing photons? And if the lines are being produced within the photosphere, the approximate surface where the medium becomes optically thin, how then are they escaping? \citep[e.g.,][]{Madau26, Perez-Gonzalez26,  Ji26, Sneppen26}. And yet, a wide array of stellar phenomena, including IIn SNe and the GE display exactly this combination of features, with the continuum being successfully modeled for several decades now as arising from an optically thick pseudo/wind/recombination-photosphere \citep[e.g.,][]{Leitherer85, Davidson87, Humphreys94, Chugai04, Dessart09, Dessart15, Owocki16}.

The key insight is that lines are produced in the wind, scattered in the wind, and escape through the wind that lies \textit{above} the pseudo-photosphere (Fig. \ref{fig:schematic}). The wind is launched close to the hot central engine/shock surface, and remains partially ionized as it expands outwards \citep[e.g.,][]{Dessart08}. The physical conditions in the wind (dense, optically thick, partially ionized) are highly conducive to processes such as collisional excitation \citep[e.g.,][]{Torralba25IFU}, Lyman pumping \citep[e.g.,][]{Kokorev25glimpsed}, Balmer scattering \citep[e.g.,][]{Chang25}, and fluorescence \citep[e.g.,][]{Tripodi25} -- these processes do not require substantial Lyman continuum photons. The $n\geq2$ levels are well-populated via collisional excitation as well as Lyman pumping -- indeed, the enshrouded eruptions display densities exactly in the regime ($\gtrsim10^{9-11}\ \rm{cm}^{-3}$) where these channels are highly efficient \citep[e.g.,][]{Drake80}. Compared to ground state atoms, $n=3-4$ atoms are much easier to collisionally excite and photoionize -- e.g., needing only $<3646$\AA\ photons that are available even in a $T_{\rm{eff}}\approx5000$ K blackbody. It is therefore not a coincidence that some of the strongest lines in LRD spectra are pumped lines such as \ion{O}{1} and \ion{Fe}{2}. The clearest signature of the dominance of collisional excitation and Balmer resonance is the \textit{median} H$\alpha$/H$\beta$ of LRDs of $\approx15$ arising from the dense gas \citep[e.g.,][]{Brooks25, Sun26, Lin26}. Finally, we note that inhomogeneities in the enveloping gas may allow hotter photons from the shock front or accreting central engine to escape into the wind, which would produce photoionization and account for high-ionization lines paralleling SN2010jl \citep[e.g.,][]{Fransson14, Tang25, Tang26}. 

We emphasize that the radiative transfer processes discussed in this section are well understood \citep[e.g.,][]{Drake80}. See e.g., the iconic SN1987A, for which \citet[][]{Xu92} note a conspicuous absence of H$\beta$ despite the strong detected H$\alpha$, paralleling the remarkable Balmer decrements of LRDs. This is explained by Balmer scattering in an optically thick medium where H$\beta$ is trapped and transformed into H$\alpha$ and Pa$\alpha$, exactly as proposed for LRDs \citep[e.g.,][]{Izotov08,Chang25, Yan25}. Further, the accompanying prominent \ion{O}{1}, seen often in both SNe and LRDs \citep[e.g.,][]{Pastorello02, Fransson02, Fransson14, Fox20}, supports this interpretation since it requires substantial optical depth in the Balmer lines for Ly-$\beta$ pumping to be efficient \citep[e.g.,][]{Grandi80}.

\section{New Approaches to LRD Central Engine Masses}
\label{sec:bhmasses}

Line profiles bearing remarkable similarities to those of LRDs are produced in IIn SNe and LBV eruptions (Fig. \ref{fig:lineprofiles}). These objects do not possess a virialized broad-line region \citep[e.g.,][]{Juodzbalis25direct, Jones25}, nor do they possess a stratified virialized broad-line region \citep[e.g.,][]{Scholtz26, Madau26stratified}, nor one sitting behind a uniform electron scattering screen \citep[e.g.,][]{ deugenio25irony, Ivey26}. Instead, the broad Balmer lines are produced and scattered in a cold, partially ionized wind (red layer in Figure \ref{fig:schematic}). Recovering an intrinsic Gaussian profile by correcting for electron scattering \citep[e.g.,][]{Rusakov26,  Kokorev25glimpsed} would yield the velocity dispersion of the cold wind, which is fundamentally different from a virial broad-line region. If this is indeed the situation in the LRDs, \textit{there is effectively nothing in the Balmer line profiles that may be connected to black hole masses via traditional calibrations}. Note that this is a fundamentally different physical scenario versus classical AGN studied using virial calibrations \citep[e.g.,][]{Greene06} -- classical AGN do not display the constellation of interlinked properties seen among LRDs.

Fresh approaches to constraining the masses of LRD central engines are needed. The analogies to stellar and supernova physics that we have drawn in this paper present a promising path that we now set out on. In what follows, we will present central engine mass estimates for the ``typical LRD" based on the sample of 116 sources selected and studied in \citet[][]{Hviding25, degraaff25pop, Sun26}. Of particular interest, the host galaxy-subtracted median stack of LRDs presented in \citet[][]{Sun26} is approximated by a $T_{\rm{eff}}=4047^{+156}_{-155}$~K (pseudo-)blackbody and has a bolometric luminosity of $L_{\rm{bol}}=10^{43.9\pm0.1}$ erg s$^{-1}$ implying an effective radius of $R_{\rm{eff}}=1321^{+168}_{-153}$ au. 

\subsection{Eddington-Limited Masses}
\label{sec:bhmass_edd}
The bolometric luminosities of LRDs are relatively well-constrained \citep[e.g.,][]{Yue24, Ananna24, Setton25, Xiao25, Casey25, Greene25}. A wide raft of models designed to reproduce various facets of LRDs invoke or predict Eddington-limited or super-Eddington accretion \citep[e.g.,][]{Loeb97, Kido25, Liu25BB, Madau26, Sneppen26, Inayoshi26specuni}. Approximate limits on central engine masses have been reported by converting the bolometric luminosity into a mass by assuming $\Gamma_{\rm{es}} = L_{\rm{bol}}/L_{\rm{edd}} = 1$ \citep[e.g.,][]{Umeda25, Greene25, Yanagisawa26, Gentile26} as follows:

\begin{align}
    M &= \frac{\kappa_{\rm{es}} L_{\rm{bol}}}{4\pi G\Gamma_{\rm{es}} c} < \frac{\kappa_{\rm{es}} L_{\rm{bol}}}{4\pi G c}.
\end{align}

Here we provide further motivation for such a mass constraint. The GE and IIn SNe display super-Eddington luminosities -- e.g., $\Gamma_{\rm{es}} \approx5-50$ \citep[e.g.,][]{Davidson16}. Adopting $\Gamma_{\rm{es}}=1$ implies $\log(M/M_{\rm{\odot}})=4.5-6.5$ for the LRD compilation in \citet[][]{degraaff25pop}. The host-subtracted stack of 98 LRDs in \citet[][]{Sun26}, which may be considered a proxy for the central engine in the typical LRD, has a luminosity commensurate with $\log(M/M_{\rm{\odot}})=5.8^{+0.1}_{-0.1}$. Higher $\Gamma_{\rm{es}}$ values of $5-50$, as in the GE and IIn SNe, bring the mass down to $\log(M/M_{\rm{\odot}})\approx4.1-5.1$ for the median LRD, and $\log(M/M_{\rm{\odot}})\approx3-6$ for the observed range in \citet[][]{degraaff25pop}.

\begin{figure}
    \centering
\includegraphics[width=0.99\linewidth]{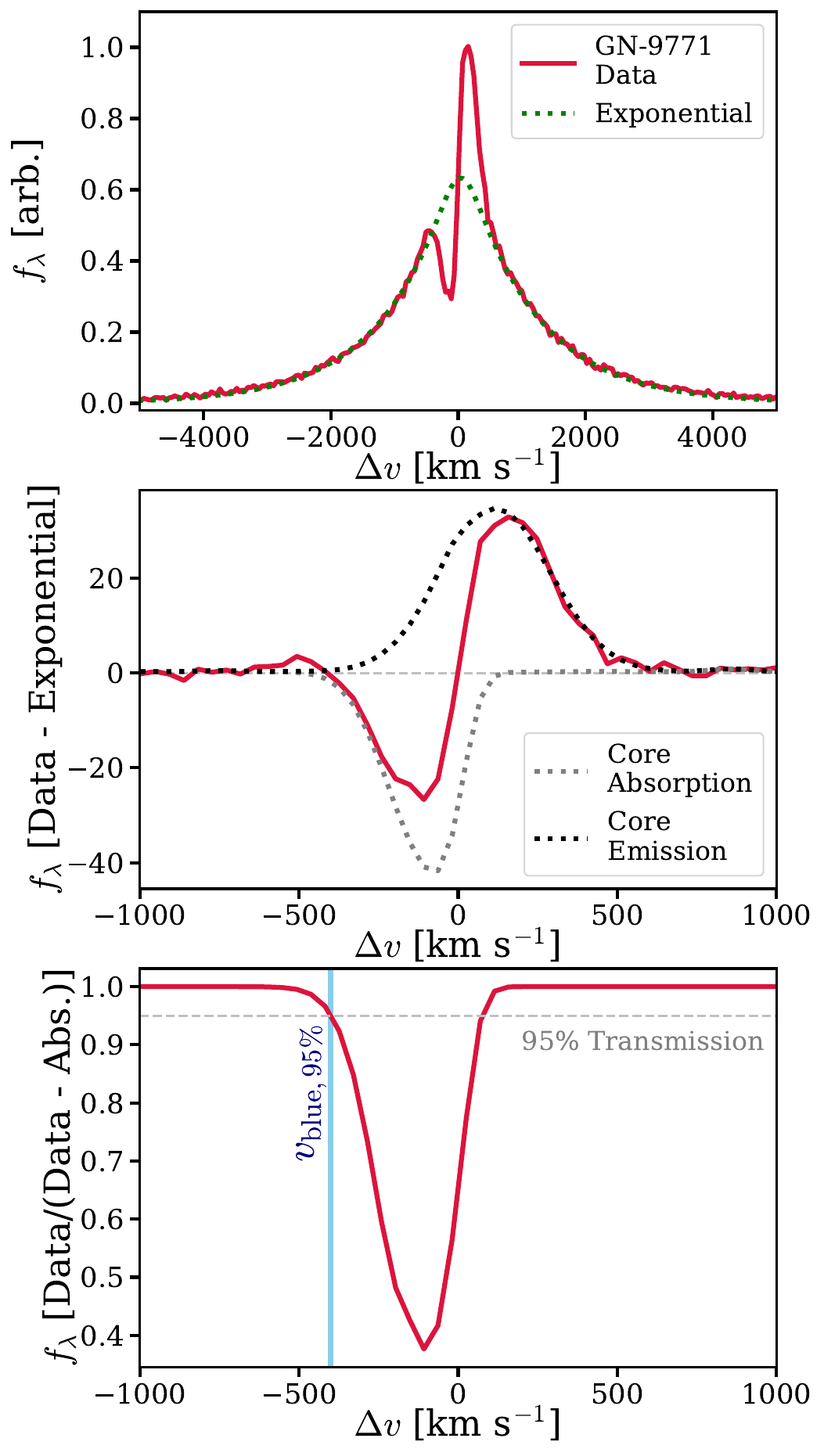}
    \caption{\textbf{Defining $v_{\rm{blue, 95\%}}$ to trace the fastest outflowing cold wind component for the escape velocity argument}. \textbf{Top:} As in \citet[][]{Matthee26}, we model the wings of the line profile as an exponential (green dashed line). \textbf{Center:} Subtracting the model for the exponential leaves us with a P-Cygni profile resembling a massive star wind. We model this profile as a combination of absorption (Voigt profile; gray) and emission (two components for Gaussian narrow emission from the host galaxy and central engine; black). \textbf{Bottom:} Using the core absorption model, we define the velocity where the transmission (i.e., Data/Data-Absorption) reaches $95\%$ as $v_{\rm{blue, 95\%}}$ (vertical blue stripe). We use this quantity as a robust proxy for the fastest outflowing cold wind component.
    }
\label{fig:v95}
\end{figure}

A simplifying assumption here is that electron scattering is the dominant source of opacity, whereas reality is likely more complex with e.g., contributions from bound-free, free-free and bound-bound opacity as well as modified force coupling between radiation and matter owing to porosity \citep[e.g.,][]{Vink11, Owocki16, Owocki17}. However, it is heartening to note that the empirical analogues, for which the central engine mass is known, return reasonable limits for $\Gamma_{\rm{es}}>1$. A key next step is to more directly demonstrate that $\Gamma_{\rm{es}}>1$ is indeed the case in the LRDs.

\subsection{An Escape Velocity Argument}
\label{sec:vesc}

The escape velocity of the cold wind, combined with the launch radius, may be used to constrain the enclosed mass as:

\begin{align}
    M = \frac{R_{\rm{esc}}v_{\rm{esc}}^{2}}{2G}.
\end{align}

The launch radius can be bounded by the radius of the pseudo-photosphere ($R_{\rm{esc}}<R_{\rm{phot}}$), which lies within the wind away from the launch surface (Fig. \ref{fig:schematic}). We can approximate $R_{\rm{phot}}$ for the pseudo-photosphere, as done in IIn and the GE \citep[e.g.,][]{Taadia13, Owocki16}, from the Stefan-Boltzmann law as:

\begin{align}
    R_{\rm{phot}} = \sqrt{\frac{L_{\rm{bol}}}{4\pi\sigma_{\rm{SB}}T_{\rm{eff}}^{4}}}.
\end{align}

For an upper limit on $v_{\rm{esc}}$, we consider the fastest outflowing material at the terminal velocity, $v_{\rm{\infty}}$. For massive star winds and eruptions, relevant to the GE and IIn SNe, $v_{\rm{\infty}}\gtrsim v_{\rm{esc}}$ with typical values of $v_{\rm{\infty}}/v_{\rm{esc}}\approx3-5$ \citep[e.g.,][]{Lamers95, MullerVink08, Hawcroft24, Schillemans26}. To practically approximate $v_{\rm{\infty}}$, in Fig. \ref{fig:v95} we define $v_{\rm{blue, 95\%}}$ to trace the blue edge of the absorber seen in LRDs. $v_{\rm{blue, 95\%}}$ is the velocity where the transmission of the absorber reaches $95\%$. We tested $\approx99\%$, but found $95\%$ to be more stable, yielding consistent results for H$\alpha$ and H$\beta$ across the entire sample. The consistency between H$\alpha$ and H$\beta$ also inspires faith that we are indeed probing $v_{\rm{\infty}}$ and not e.g., a radial density gradient which would produce different cutoffs for these lines. We estimate an upper-limit on the mass of the central engine as:
\begin{align}
    M < \frac{R_{\rm{phot}}v_{\rm{blue,95\%}}^{2}}{2G}.
\end{align}

Constraining $v_{\rm{blue, 95\%}}$ requires high-resolution high-SNR spectra. \citet[][]{Matthee26} recently compiled and homogeneously analyzed such a sample of 18 sources that capture almost the entire spread in luminosity among known LRDs ($\log(L_{\rm{bol}}/L_{\rm{\odot}})\approx9-10.5$; see Fig. \ref{fig:hrd}). Of these 18 sources, 11 have well-detected absorbers for which we perform the analysis shown in Fig. \ref{fig:v95}. 

We find a median $v_{\rm{blue, 95\%}}=-443^{+43}_{-126}$ km s$^{-1}$ for this sample (see Appendix for parameters for individual sources). Combined with the typical $R_{\rm{phot}}\approx1300$ au for LRDs (Fig. \ref{fig:hrd}; \citealt{Sun26}) we find $\log(M_{\rm{}}/M_{\rm{\odot}})<5.1$, which translates to $L_{\rm{bol}}/L_{\rm{edd}}>5$. Despite our conservative assumptions, this is a remarkably illuminating limit -- we are able to infer that the LRDs are likely super-Eddington systems. 

As a sanity check, we apply this argument to the GE, where the true enclosed mass is constrained via the orbit of the $\eta$ Car binary as well as detailed studies of the remnant, the Homunculus nebula \citep[e.g.,][]{Smith03, Morris17, Strawn23}. For an $R_{\rm{phot}}\approx30$~au (Fig. \ref{fig:hrd}) and $v_{\rm{blue,95\%}}\approx300$ km s$^{-1}$ (Fig. \ref{fig:lineprofiles}) we find an upper limit of $M_{\rm{}}\lesssim1500\  M_{\rm{\odot}}$. The true enclosed mass, including the mass lost in the GE, is $\approx150$ $M_{\rm{\odot}}$. That is, the derived enclosed mass is indeed a conservative upper limit and lies $\approx10\times$ higher than the true value. If the mass of LRD central engines is similarly overestimated, this yields $\log(M_{\rm{}}/M_{\rm{\odot}})\approx4$ for the typical LRD.

\begin{deluxetable*}{lccc}
\tabletypesize{\footnotesize}
\tablecaption{Summary of LRD Central Engine Mass Estimates\label{tab:mass}}
\tablehead{
\colhead{} & \multicolumn{2}{c}{Typical LRD} & \colhead{Sample Range} \\
\colhead{Argument} & \colhead{Conservative Limit} & \colhead{Best Guess} & \colhead{Best Guess} }
\startdata
Eddington-Limited (\S\ref{sec:bhmass_edd}) & $<5.8$ & $4.1-5.1$ & $2.8-5.8$ \\
Escape Velocity (\S\ref{sec:vesc}) & $<5.1$ & $4.1$ & $3.0-5.3$ \\
Variability (\S\ref{sec:bhmass_var}) & $<5.5$ & $4.5$ & $3.3-5.6$\\
Surface Gravity (\S\ref{sec:bhmass_logg}) & $<5.5$ & $\cdots$ & $\cdots$
\enddata
\tablecomments{Masses reported as $\log(M/M_{\rm{\odot}})$. The ``typical LRD" is the median host-subtracted stack from \citet[][]{Sun26} with the following parameters: $\log(L_{\rm{bol}}/\rm{erg}\  \rm{s^{-1}})=43.9^{+0.1}_{-0.1}$, $T_{\rm{eff}}=4047^{+156}_{-155}$ K, $R_{\rm{eff}}=1321^{+168}_{-153}$ au. The sample range is based on \citet[][]{degraaff25pop} by bracketing various quantities of interest (e.g., $R_{\rm{phot}}$, $T_{\rm{eff}}$, $L_{\rm{bol}}$) based on their $5^{\rm{th}}$ and $95^{\rm{th}}$ percentiles. See text for details on the various arguments.}
\end{deluxetable*}

\begin{figure*}
    \centering \includegraphics[width=0.7\linewidth]{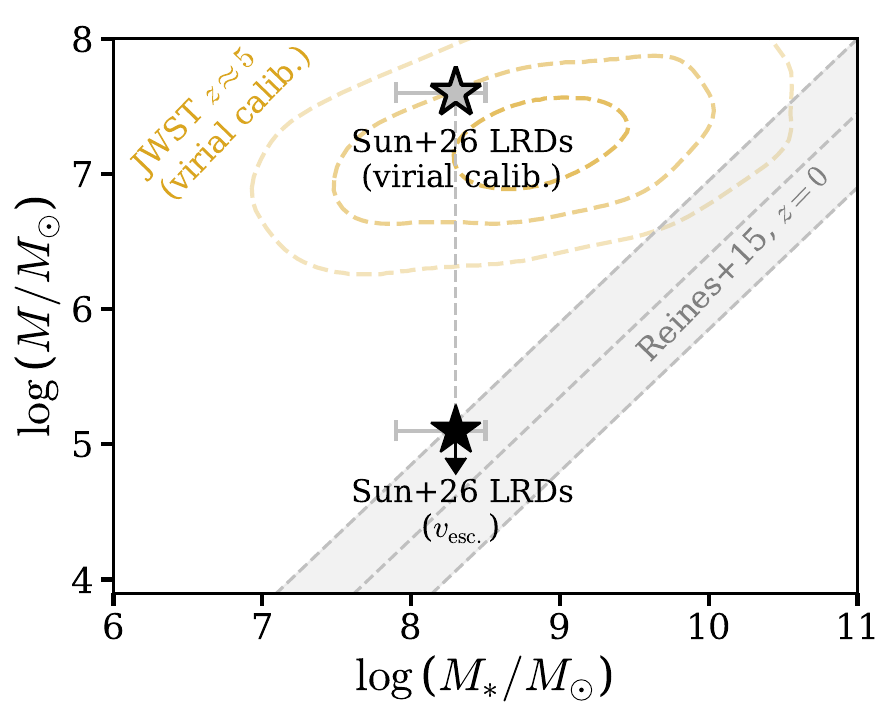} \caption{\textbf{``Overmassive" no more -- central engine mass vs. stellar mass for the typical LRD}. We display parameters estimated for the stack of 98 LRDs from \citet[][]{Sun26} (black stars) that were decomposed into a host galaxy and a central engine. The stellar mass is based on SED fitting and validated by clustering measurements \citep[e.g.,][]{Matthee25LRDclustering, Lin25clustering, Pizzati25}. Our inferred central engine mass for this stack ($<10^{5.1} M_{\rm{\odot}}$) is consistent across four lines of argument outlined in \S\ref{sec:bhmasses} and places LRDs in the intermediate mass regime. Here we show the upper limit based on $v_{\rm{esc}}$, the most constraining of our four arguments in \S\ref{sec:bhmasses} (see Table \ref{tab:mass}). The LRDs lie only $<0.4$ dex above, and within the scatter, of the $z=0$ relationship between stellar and black hole mass \citep[e.g.,][]{RV15}. This is in sharp contrast with widespread JWST estimates that report ``overmassive" black holes lying orders of magnitude off the relation by assuming $z=0$ virial calibrations that are likely inapplicable to LRDs and high-$z$ AGN (gold contours; compilation from \citealt{Li25}). For illustration, we apply the \citet[][]{Reines13} virial calibrations to the \citet[][]{Sun26} stack ($L_{\rm{H\alpha}}=10^{42.8}$ erg s$^{-1}$) assuming a typical LRD Gaussian FWHM(H$\alpha$)=2100 km s$^{-1}$ \citep[][]{Matthee24}, with the resulting ``overmassive" black hole shown as a gray star.}
    \label{fig:mbhmstar}
\end{figure*}

A complication for this argument, which we ignore here in this first attempt, is that LRD winds may have a more complex structure. This is evidenced by the absorption extending to redshifted velocities reported in $\approx15-20\%$ of LRDs with high-resolution observations \citep[e.g.,][]{Matthee26, Davis26, Lin26}. This may be due to a rotational component in LRD central engines and their possible progenitors \citep{Torralba26panbhstar} -- e.g., rotation may be a generic feature of supermassive stars and the accretion disks around them \citep[e.g.,][]{Haemmerle21, Zwick26}, while material accreting onto a black hole likely also carries angular momentum \citep[e.g.,][]{Hopkins24}. With rotation, even a strong outflow can manifest as redshifted absorption due to projection effects. Another possibility is that accretion or inflows onto the central engine from e.g., the surrounding ISM \citep[e.g.,][]{Kido25} contribute to the redshifted absorption. Practically speaking, these considerations are of concern if $v_{\rm{blue, 95\%}}$ is underestimating $v_{\rm{\infty}}$ (an overestimate is acceptable for our upper limit), which would be the case if we were probing only the highest optical depth regions of the outflow, and not the full outflow. Note that based on the sanity check against GE, and models of GE-like outbursts \citep[e.g.,][]{Schillemans26}, $v_{\rm{\infty}}\gg v_{\rm{esc}}$ at $R_{\rm{phot}}$, and so our conservative upper limit leaves significant room to capture the true mass. Building detailed wind models and constraining them against e.g., multiple species with different line formation depths (e.g., \ion{He}{1} vs. H$\alpha$ vs. H$\beta$) and forward modeling population wind profiles to marginalize over geometry are promising avenues for future work.

\subsection{Super-Eddington Central Engines from Variability}
\label{sec:bhmass_var}

Earlier we discussed the diffusion timescale ($t_{\rm{diff}}$,\S\ref{sec:diffusion}) -- the characteristic timescale over which changes interior to the pseudo-photosphere are smeared. Here we consider the dynamical time ($t_{\rm{dyn}}$) -- the characteristic time over which the extended pseudo-photosphere responds mechanically to gravity. The same constraints on variability discussed in \S\ref{sec:diffusion} place a lower limit on $t_{\rm{dyn}}$, i.e., the typical LRD's $t_{\rm{dyn}}=R_{\rm{esc}}/v_{\rm{esc}}$ appears to be $\gtrsim10$ years \citep[e.g.,][]{Liu26TWINKLE,Burke25,Lin25,Zhang25cepheid, Furtak25,Park26}, with some hints of broadband photometric variability showing up over a 100+ year baseline enabled by gravitational lensing in one object \citep[][]{Zhang25cepheid}. Conservatively adopting $t_{\rm{dyn}}>10$ years and $R_{\rm{esc}}<R_{\rm{phot}}$ yields an upper limit on the enclosed mass as follows: 

\begin{align}
     M_{\rm{}} &  <  \frac{R_{\rm{phot}}^{3}}{2Gt_{\rm{dyn}}^{2}}.
\end{align}

This implies $M_{\rm{}}<10^{5.5} M_{\rm{\odot}}$. Taking the putative Cepheid-like pulsation period from \citet[][]{Zhang25cepheid} as typical ($t_{\rm{dyn}}=30$ yrs) would yield $M_{\rm{}}\approx10^{4.5} M_{\rm{\odot}}$. Indeed, \citet{Liu26TWINKLE, Secunda26} argue that current variability constraints on ensembles of LRDs rule out standard sub-Eddington models, favoring low black hole masses and super-Eddington accretion. 

\subsection{Surface Gravity of the Recombination Photosphere}
\label{sec:bhmass_logg}

The pseudo-photosphere to some extent may be approximated as a very low-density stellar atmosphere \citep[e.g.,][]{Liu26TLUSTY}. Though note the key differences discussed earlier -- wind photospheres have a far more extended spectrum formation region with a density/temperature gradient \citep[][]{Davidson16}, and of course, are in motion. ``Stellar" parameters may be constrained from absorption features such as the Calcium Triplet (CaT) and the continuum shape \citep[e.g.,][]{Lin25, Torralba26panbhstar}. These parameters, in particular the surface gravity, $\log(g)$, translate to constraints on the mass. For a medium with radially increasing outward velocity, as is the case for a wind, the apparent $\log(g)$ from the spectrum is an upper limit on the true surface gravity \citep[][]{Liu26TLUSTY} and so:

\begin{align}
M_{\rm{}} &< \frac{gR^{2}_{\rm{phot}}}{G}.
\end{align}

Adopting a surface gravity of $\log(g)\approx-1$ suggested by comparing the CaT EW observed in the local $z\approx0$ LRD, ``The Egg" \citep[e.g.,][]{Lin25, Ji25lol}, with Milky Way stars \citep[][]{Lin25}, we infer $M_{\rm{}}<10^{5.5} M_{\rm{\odot}}$ for $R_{\rm{phot}}=1300$ au in the typical LRD \citep[][]{Sun26}. Using low-density stellar atmosphere models, \citet{Liu26TLUSTY} advocate for the degenerate $\log(g)\approx-3$ solution that produces a similar CaT EW but a bluer NIR continuum in better agreement with data. Our radially accelerating wind picture would then suggest an even tighter mass constraint of $M_{\rm{}}<10^{3.5} M_{\rm{\odot}}$. A key caveat here is that the Egg is an extremely low-luminosity object with respect to higher redshift LRDs ($\approx5\times10^{9} L_{\rm{\odot}}$, \citealt{Lin25}; compare with Fig. \ref{fig:hrd}), and its $\log(g)$ might not be representative. Further, the NIR is likely being altered by free-free and dust emission in the wind at large radii, which are typically seen in massive star winds and Type IIn SNe (see Fig. \ref{fig:freefree}) but are not self-consistently included in the \citet{Liu26TLUSTY} hydrostatic models. Refining this upper limit requires detailed wind spectral modeling that we defer to future work.

Hearteningly, the four independent lines of argument presented here, inspired by the quasi-stellar physics of LRDs, all paint a consistent picture --  $\lesssim10^{5} M_{\rm{\odot}}$, intermediate mass, super-Eddington central engines. We summarize these estimates in Table \ref{tab:mass}. With these reduced BH masses, LRDs are consistent with the $z\approx0$ relationship between stellar and black hole masses (Fig. \ref{fig:mbhmstar}). This resolves the fundamental tension that the $z\approx5-9$ LRD black hole mass density, when virial relations are applied verbatim, may already account for almost the entire $z=0$ black hole mass density \citep[e.g.,][]{InayoshiIchikawa24, Umeda25, Luberto25, Shen26}.

\section{Discussion}
\label{sec:discussion}

\subsection{What lies at the heart of LRDs?}

Until this point, we have remained agnostic. All the results presented here do not assume any central engine. A key implication of the enshrouded eruption analogy is that it is extremely challenging to gather any \textit{direct} information about what exactly lies within the dense gas envelope. To draw a parallel from the SNe literature, some apparent ``IIn SNe" are not supernovae at all, but are instead GE-like outbursts (``supernova impostors"; e.g., \citealt{vandyk00,Smith10, Humphreys12}). A dramatic example is ``SN2009ip" that in reality was a GE-like eruption -- only to then explode as a IIn SN in 2012 \citep[e.g.,][]{Mauerhan13}. The ambiguous class of  ``IIn/Ia-CSM SNe" -- where the nature of the explosion, thermonuclear (Ia) or core-collapse (II/Ib/Ic), is unknown -- further underscores the difficulty of seeing past the pseudo-photosphere \citep[e.g.,][]{Inserra16, Jerkstrand20, Sharma23}. Long-term monitoring is what unmasks the SNe impostors, but the timescales relevant for LRDs (\S\ref{sec:diffusion}, \S\ref{sec:bhmass_var}) are more challenging to observe. 

Keeping this ambiguity inherent to the physical setup in mind, based on indirect arguments and analogies we discuss two leading candidates for LRD central engines -- supermassive stars and intermediate mass black holes\footnote{The unknowability and Schr\"odinger's cat-like state of what lies within the dense gas envelope gives added new meaning to ``black hole star" as a superposition of the IMBH/SMS possibilities.}.

\subsubsection{Supermassive Star Central Engines}
\label{sec:sms}

Supermassive stars (SMS; $10^{3-5} M_{\rm{\odot}}$) may play a role in LRD formation \citep[e.g.,][]{Nandal25, Chisholm26, Martins26, Nandal26, Zwick26}. LRDs have a propensity for strong Nitrogen emission \citep[e.g.,][]{Isobe25, Morel25}, implying extreme [N/O] and [N/C] abundances that are typically found in the Milky Way among stars born in dense, massive clusters \citep[e.g.,][]{Belokurov23, Belokurov24, Naidu25z14,Schaerer25GCs, Ji25Nitrogen}. Dense star clusters are exactly where SMS are predicted to form via runaway collisions and produce such peculiar abundances \citep[e.g.,][]{Denissenkov14, Gieles18, Charbonnel23, Nandal25Nitrogen, Rantala26, Williams26}. Further, the age distribution of metal-poor Milky Way globular clusters -- possible birth sites of SMS -- may mirror the redshift evolution of the LRD number density \citep[][]{Chisholm26}. \citet[][]{Sun26} demonstrate LRD hosts to have undergone recent starbursts, possibly linking the formation (or activation) of LRD central engines to very recent star-formation, which at high-redshift proceeds in the form of dense cluster formation \citep[e.g.,][]{Fujimoto24, Adamo24,Claeyssens25}. Finally, the LRD luminosity function has a sharp cutoff at the bright-end \citep[e.g.,][]{Ma25countingLRDs, Weibel26}, which may correspond to the general relativistic instability limit for the maximum mass of an SMS ($\approx10^{5-6} M_{\rm{\odot}}$; e.g., \citealt{Woods17, Nandal24gr, Saio24gr}) provided it has an $L_{\rm{bol}}/L_{\rm{edd}}$ comparable to the GE and IIn SNe.

Against this backdrop, given the connections to massive star phenomena made in this work (e.g., the GE arising from a $\approx10^{2} M_{\rm{\odot}}$ star), it is natural to conjecture that the formation of LRDs may be linked to even more massive stars ($\gtrsim10^{3} M_{\rm{\odot}}$). We envision that an SMS eruption that enshrouds the star would be followed by a succession of increasingly violent eruptions that could power the LRD phase via CSM interaction. Each discrete eruption -- of which there may be several dozens \citep[e.g.,][]{Nandal26} -- may sustain an LRD-like spectrum for decades to centuries (\S\ref{sec:diffusion}) over the $\approx1-2$ Myr life of the SMS. Enshrouding the central engine in this manner elegantly sidesteps the issue of accreting external gas and robbing it of angular momentum for it to settle into a fully enclosing (covering fraction approaching $\approx100\%$) gas envelope \citep[e.g.,][]{Yanagisawa4pi}. 

Other major challenges for SMS as LRD progenitors are alleviated by our results. General relativistic (GR) instabilities set a limit of $\approx10^{5-6} M_{\rm{\odot}}$ for SMS survival under standard assumptions \citep[e.g.,][]{Woods17, Nandal24gr, Saio24gr}. The reduced central engine masses we find (Table \ref{tab:mass}) are now compatible with this limit. Furthermore, the CSM interaction mechanism from a succession of winds can elevate the luminosity of $<10^{5} M_{\rm{\odot}}$ SMS ($\approx 10^{9.5} L_{\rm{\odot}}$ at the Eddington limit) to encompass the entire range for observed LRDs ($\approx10^{9}-10^{11.5} L_{\rm{\odot}}$; see Fig. \ref{fig:hrd}). Further, these lower central engine masses reconcile upper limits on BH masses in local GCs \citep[e.g.,][]{Baumgardt17} with scenarios where SMS forming in GCs give rise to LRDs \citep[e.g.,][]{Chisholm26}.

A major remaining challenge is that the short SMS lifetimes of $\lesssim1-2$ Myr, even under the most optimistic assumptions, including continuous external accretion \citep[e.g.,][]{Nandal26SMSlifetimes} are in tension with LRD duty cycle constraints. The typical LRD host spends $\approx10-20$ Myrs as an LRD \citep[e.g.,][]{Sun26}. Further, note that the kind of dense, optically thick winds required to match the properties of LRDs might appear only towards the very end of the SMS life just before it collapses into an IMBH (in the last $\approx0.1$ Myr; e.g., \citealt{Nandal26}), though these models are still in the very early stages of development \citep[e.g.,][]{Martins26}. This tension may be somewhat alleviated if several SMS form within the same galaxy in distinct star clusters, as hinted by a $z\approx7$ galaxy with multiple LRDs \citep[][]{Yanagisawa26} -- together, they may account for the integrated lifetime. This is not unreasonable, given the multiplicity of GCs observed in galaxies of mass comparable to typical LRD hosts ($\approx10^{8} M_{\rm{\odot}}$; e.g., \citealt{Matthee25LRDclustering,Sun26}). For instance, arguably the best-studied high-redshift ($z\approx2$) galaxy with mass comparable to LRD hosts ($\approx10^{8} M_{\rm{\odot}}$; \citealt{Matthee25LRDclustering, Lin25clustering, Sun26}), the \textit{Gaia}-Sausage Enceladus \citep[e.g.,][]{Helmi18, Belokurov18, Naidu21}, may have hosted an ensemble of $\approx30$ GCs \citep[e.g.,][]{Myeong18, Massari19, Kruijssen20}. Indeed, the average number of GCs hosted by an LRD host galaxy ($M_{\rm{\star}}\approx10^{8} M_{\rm{\odot}}$, $M_{\rm{halo}}\approx10^{11} M_{\rm{\odot}}$ at $z\approx5$; \citealt{Lin25clustering, Matthee25LRDclustering, Sun26}) based on simple analytical models of GC formation is $\approx10-20$ \citep[e.g.,][]{Elbadry19}.

We conclude that if not for the central engines of LRDs themselves, SMS are excellent candidates for LRD progenitors (i.e., the ``Pre-Eruption" LRD phase in Fig. \ref{fig:schematic}).

\subsubsection{All Roads Lead to Super-Eddington IMBHs}
\label{sec:imbhs}

Inspired by the massive black holes ($\approx10^{8-9} M_{\rm{\odot}}$) already in place in the first billion years \citep[e.g.,][]{Fan01, Banados18, Yang20poni, Wang21, Fan23}, it has long been anticipated that their rapid rise may require spectacular physical phenomena including e.g., supermassive stars, hyper-Eddington accretion, and dense gas shrouds \citep[e.g.,][]{Rees84, Begelman06, Alexander14, Wise19, Inayoshi20, Volonteri21, Natarajan21}. 

The terminal state of an SMS (or almost any alternative for what lies within the gas envelope) is likely to be an IMBH \citep[e.g.,][]{Umeda16, Woods17, Nandal26}. Indeed, this is supported by our central engine mass constraints in \S\ref{sec:bhmasses}. And so LRDs either already represent the long-awaited discovery of early Universe super-Eddington IMBHs \citep[e.g.,][]{Greene20, Inayoshi20}, or are a few Myrs away from turning into these IMBHs. Further, if these IMBHs are descended directly from SMS -- as suggested by the case for SMS in \S\ref{sec:sms} -- LRDs represent the heavy seeds envisioned by numerous massive black hole formation channels \citep[e.g.,][]{Lodato07, Begelman08, Volonteri10}. 

In the context of our proposed picture (\S\ref{sec:physics}, Fig. \ref{fig:schematic}), IMBHs are long-lived, can drive strong winds (in e.g., the form of super-Eddington outflows; e.g., \citealt{Liu25}), and can supplement or act in place of the radiative energy from CSM interaction with accretion energy. We again emphasize that effectively all signatures that have been interpreted to arise from an AGN in LRDs -- e.g., broad Balmer lines, high bolometric luminosity -- can be explained as a product of wind and CSM interaction physics. However, it is interesting to note that there exist a handful of sources that share (some) features of LRDs (e.g., V-shaped or blackbody-like SEDs), but are X-ray bright \citep[e.g.,][]{Fu26FORGES, Hviding26, Zhong26bbqsos}. If indeed connected to LRDs, these may be transitional objects where the gas envelopes are dispersing to reveal an accreting IMBH. As such, they currently represent the only relatively direct line of evidence for the LRDs' IMBH nature.

\subsection{What follows the LRD phase? Dust Production and the Beginnings of Classical AGN}
\label{sec:dust}

Enshrouded eruptions are highly effective at producing dust \citep[e.g.,][]{Fox11, Maeda13, Smith13, Gall14, Morris17, Shahbandeh25}. LRDs may similarly be highly efficient dust factories. Dust may form in dense winds when the outflow cools and becomes sufficiently shielded from the central radiation field, allowing refractory atoms and molecules to nucleate into seed grains that rapidly grow before the gas becomes too dilute. Indeed, hints of such dust are seen most clearly among a subset of $z\approx0$ LRDs in the infrared \citep[e.g.,][]{Ji25lol, Lin25, Park26, Lin26}, whose infrared excesses above the photospheric continuum may be attributed to freshly forming dust (e.g., Carbon and Silicates; see Fig. \ref{fig:freefree}). Further, molecules such as water are beginning to be detected in LRDs \citep[e.g.,][]{Wang26}, which show that the optimal conditions for dust-formation such as cold layers with $T<3000$ K are indeed found in these sources. LRDs may therefore be an abundant, previously unconsidered source of dust in the very early Universe -- an epoch where signatures of dust are routinely detected, but whose source remains unclear \citep[e.g.,][]{Witstok23, Shivaei25, Markov25, Algera26}. We defer detailed exploration of the implications for high-redshift dust physics to Ashall et al. in prep. 

Here we note an intriguing implication for massive black hole evolution -- the extent of the pseudo-photosphere ($\approx1300$ au) already rivals that of broad-line regions in low-luminosity AGN \citep[e.g.,][]{Bentz13}. With dust being produced prolifically, all the ingredients are in place to form the entire structure of a classical broad-line region such as a gas disk with a dusty torus. Given this, it seems plausible that at the end of the enshrouded eruption phase, LRDs are poised to transform into the familiar AGN of the low-$z$ Universe \citep[e.g.,][]{Urry95}.

\section{Summary}

We argue that LRDs share numerous empirical features with enshrouded eruptions of massive stars. This is rather remarkable, since many of these LRD features are exceptionally rare across all galaxies and AGN -- their collective co-occurrence in a well-understood set of objects promises rich insights. 

\vspace{0.3cm}
\noindent $\bullet$ These shared features described in \S\ref{sec:empirical} include:  

\begin{itemize}

\item \textit{Blackbody-like Continuum} from a ``pseudo-photosphere" in the dense, outflowing ``slow wind" enshrouding the central engine. The $T_{\rm{eff}}\approx5000$ K is set by the ionization-dependent opacity of a hydrogen-rich plasma. [Fig. \ref{fig:hrd}] 

\item \textit{Molecular signatures (e.g., H$_{\rm{2}}$O, CO, CN}) that are imprinted by $T<3000-4000$ K layers in the wind. 

\item \textit{Optical luminosity comparable to bolometric luminosity} -- X-rays and FUV photons are absorbed and reprocessed into optical and IR light in the dense envelope, in sharp contrast to typical AGN.

\item \textit{MIR excesses above the pseudo-photospheric continuum} produced by free-free emission from the wind and dust emission from newly forming dust. [Fig. \ref{fig:freefree}, \S\ref{sec:dust}]

\item \textit{P-Cygni Profile Cores + Electron Scattered Wings} in e.g., \ion{He}{1}, H$\alpha$, and H$\beta$ that form and are then scattered in the dense, partially ionized, optically thick wind. The extent and kinematics of electron scattering and Balmer absorption varies with line formation optical depth. [Fig. \ref{fig:lineprofiles}, \ref{fig:hasubtle}, \ref{fig:electronscattering}]

\item \textit{Extreme Balmer decrements (H$\alpha$/H$\beta>10$) from gas, not dust}. Optically thick conditions facilitate collisional excitation as well as Balmer scattering, which converts H$\beta$ into Pa$\alpha$ and H$\alpha$. [Fig. \ref{fig:hasubtle}]

\item \textit{Strong \ion{O}{1} and \ion{Fe}{2} transitions} produced by Lyman-$\alpha,\beta$ pumping and collisional excitation that is highly efficient in dense gas. [Fig. \ref{fig:fullspec}]

\item \textit{Dearth of [\ion{O}{3}] 4960, 5008\AA}, reflecting gas densities above the critical density (e.g., $\gtrsim10^{9}$ cm$^{-3}$) and weak ionization. [Fig. \ref{fig:fullspec}, \ref{fig:snphases}]

\item \textit{The rare tail of blue, low column density LRDs} may be analogous to short-lived phases of IIn SNe, such as shock breakout, when hard ionizing photons traverse the cold wind to produce \ion{He}{2} emission alongside electron scattered Balmer lines with no absorption. [Fig. \ref{fig:snphases}]

\end{itemize}

\noindent $\bullet$ Inspired by these striking similarities, we sketch a two-stage model for LRDs adapting the relatively well-understood physics of the GE and IIn SNe. [Fig. \ref{fig:schematic}, \S\ref{sec:physics}]

\begin{itemize}
    \item \textit{Stage I -- The Pre-LRD Phase:} The central engine drives a wind of highly dense, optically thick material. A hydrogen recombination pseudo-photosphere forms \textit{within} the wind, thoroughly obscuring the hot surface of the central engine. Possible sources for the winds include e.g., supermassive star eruptions, outflows from accretion disks.
    
    \item \textit{Stage II -- The LRD Phase:} The central engine drives fast winds into the pre-existing envelope. Kinetic energy from the resulting shocks is converted highly efficiently into radiation ($\approx50\%$ efficiency vs. $<1\%$ in typical SNe). This explains the LRDs' immense luminosity with physics similar to interaction transients, which rank among the most luminous transients. Accretion radiation may complement or act in lieu of CSM interaction luminosity.
    
    \item The photon diffusion timescale of LRDs is on the order of decades to centuries, explaining the lack of variability over shorter monitoring baselines despite this transient-like physics. [\S\ref{sec:diffusion}]
    
    \item We estimate a mass-loss rate of $\approx0.3 M_{\rm{\odot}}$ yr$^{-1}$ and a fast wind velocity of $\approx8000$ km s$^{-1}$ that crashes into a slow wind of $\approx500$ km s$^{-1}$. The wind velocities are comparable to IIn SNe and the GE, while the mass-loss rates are $\approx10-100\times$ higher, consistent with a more massive central engine. [\S\ref{sec:windparams}]
    
    \item  The co-existence of a pseudo-photospheric continuum with strong emission lines is a natural outcome of this picture. Lines form above the pseudo-photosphere, excited by collisions, resonance, and fluorescence in dense, optically thick gas. [\S\ref{sec:coexist}]
    
\end{itemize}

\noindent $\bullet$ The enshrouded physics of LRDs implies that Balmer lines are formed and scattered in the cold wind, and are therefore not tracing a virial broad line region. We present four arguments for central engine masses --

\begin{itemize}
    \item Assuming LRDs are super-Eddington by analogy to the GE and IIn SNe ($L_{\rm{bol}}/L_{\rm{edd}}\approx5-50$). [\S\ref{sec:bhmass_edd}]
    \item Using the fastest outflowing absorbing gas in an ``escape velocity" argument. [\S\ref{sec:vesc}]
    \item Translating the lack of variability into a dynamical timescale of the pseudo-photosphere. [\S\ref{sec:bhmass_var}]
    \item Converting the low surface gravity of the pseudo-photosphere into an enclosed mass. [\S\ref{sec:bhmass_logg}]
\end{itemize}

All four arguments consistently imply intermediate mass, super-Eddington central engines with $M\lesssim10^{5} M_{\rm{\odot}}$ and $L_{\rm{bol}}/L_{\rm{edd}}\gtrsim5$ for the typical LRD, with a population range of $\approx10^{3-6} M_{\rm{\odot}}$. There may be no ``overmassive black holes", with LRDs falling within the scatter of the local $M_{\rm{\star}}-M_{\rm{BH}}$ relationship. [Table \ref{tab:mass}, \S\ref{sec:bhmasses}, Fig. \ref{fig:mbhmstar}]

\vspace{0.3cm}
\noindent $\bullet$ These lower central engine masses, the numerous parallels between LRDs and massive star phenomena, and the route to super-Eddington luminosity via wind interaction highlighted in this paper strengthen the viability of supermassive stars as LRD progenitors. [\S\ref{sec:sms}]

\vspace{0.3cm}
\noindent $\bullet$ All signatures that have been interpreted to arise from an AGN in LRDs (e.g., broad lines, high luminosity) may be explained via wind interaction. However, circumstantial evidence (e.g., X-ray detected LRD-like objects, lifetimes, that all SMS end up as BHs) suggests IMBHs may be the protagonists of the LRD phase. [\S\ref{sec:imbhs}]

\vspace{0.3cm}
\noindent $\bullet$ Like enshrouded eruptions, LRDs have the perfect combination of cold, shielded dense gas to be prolific producers of dust, representing a new channel for the early Universe dust budget. This dust production may mark the beginnings of classical AGN with e.g., dusty tori. [\S\ref{sec:dust}]

\vspace{0.3cm}

It is a special thing in the study of the Universe to be confronted by an entirely new category of astrophysical object. Just as the Boorong people were transfixed by $\eta$ Carinae, and just as the astronomers of the 50s were seized by quasi-stellar objects, we find ourselves again in this special place with the Little Red Dots.

\section*{Acknowledgments}

RPN dedicates this paper to the memory of A. Neil Pappalardo (1942-2026), whose infinite enthusiasm for the Universe lives on in his Pappalardo Fellows' pursuits. 

We acknowledge illuminating conversations with Dave Coulter, Taya Govreen-Segal, Jenny Greene, Calum Hawcroft and Kohei Inayoshi. We thank Daichi Hiramatsu for advice on Fig. \ref{fig:hrd} featuring IIn SNe from \citet[][]{Hiramatsu24}. We thank Erkki Kankare for sharing data on SN2009kn and Jesper Sollerman for sharing data on SN1994W. We are grateful to the organizers of the ``Towards the Full Bloom of First Star, Galaxy,
and Black Hole Formation Exploration" conference in Tokyo (March 2026) and the ``Charting Cosmic Dawn" conference in Copenhagen (April 2026) for creating a stimulating environment for discussions of LRDs.

RPN acknowledges the generous support of Neil and Jane Pappalardo via the MIT Pappalardo Fellowship in Physics. Support for this work was provided by NASA through the NASA Hubble Fellowship grant HST-HF2-51515.001-A awarded by the Space Telescope Science Institute, which is operated by the Association of Universities for Research in Astronomy, Incorporated, under NASA contract NAS5-26555. RPN, WQS, and ZL acknowledge funding from JWST-GO-3516, JWST-GO-5224, and JWST-GO-7404. JM and AT acknowledge funding from the European Union (ERC, AGENTS,  101076224). Views and opinions expressed are, however, those of the authors only and do not necessarily reflect those of the European Union or the European Research Council. Neither the European Union nor the granting authority can be held responsible for them. This research was supported by the International Space Science Institute (ISSI) in Bern, through ISSI International Team 25-659, ``Little Red Dots, Big Open Questions: Unraveling the Mystery of the James Webb Space Telescope's Most Debated Discovery" co-led by RPN and MX.  C.A. acknowledges support from NASA grants JWST-GO-02114,
JWST-GO-02122, JWST-GO-04522, JWST-GO-04217, JWST-GO-04436,
JWST-GO-03726, JWST-GO-05057, JWST-GO-05290, JWST-GO-06023,
JWST-GO-06677, JWST-GO-06213, JWST-GO-06583. Support for
programs \#2114, \#2122, \#3516, \#3726, \#4217, \#4436, \#4522,  \#5057, \#5224,
\#6023, \#6213, \#6583, \#6677, and \#7404
were provided by NASA through a grant from the Space Telescope Science
Institute, which is operated by the Association of Universities for Research in
Astronomy, Inc., under NASA contract NAS 5-03127.
This work has received funding from the Swiss State Secretariat for Education, Research and Innovation (SERI) under contract number MB22.00072, as well as from the Swiss National Science Foundation (SNSF) through project grants 200020\_207349 and 2000-1-243073. 
AdG acknowledges support from a Clay Fellowship awarded by the Smithsonian Astrophysical Observatory. REH acknowledges support by the German Aerospace Center (DLR) and the Federal Ministry for Economic Affairs and Energy (BMWi) through program 50OR2403 `RUBIES'. D.O.J. acknowledges support from NSF grants AST-2407632, AST-2429450, and AST-2510993, NASA grants 80NSSC24M0023 and 80NSSC24K0353, and HST/JWST grants HST-GO-17128.028 and JWST-GO-05324.031, awarded by the Space Telescope Science Institute (STScI), which is operated by the Association of Universities for Research in Astronomy, Inc., for NASA, under contract NAS5-26555.  This work is also funded by the Gordon and Betty Moore Foundation through Grant GBMF13900 to D.O.J.
Some of the data products presented herein were retrieved from the Dawn JWST Archive (DJA). DJA is an initiative of the Cosmic Dawn Center (DAWN), which is funded by the Danish National Research Foundation under grant DNRF140. This work is based on observations made with the NASA/ESA/CSA James Webb Space Telescope. The data were obtained from the
Mikulski Archive for Space Telescopes at the Space Telescope Science Institute, which is operated by the Association of Universities for Research in Astronomy, Inc., under NASA contract NAS 5-03127
for JWST. Support for programs
\#3516, \#5224, \#5664, \#7404 was provided by NASA through grants from the Space
Telescope Science Institute, which is operated by the Association of
Universities for Research in Astronomy, Inc., under NASA contract
NAS 5-03127. 

Software used in developing this work includes: \texttt{claude} \citep[][]{Anthropic2025ClaudeOpus47}, \texttt{matplotlib} \citep{matplotlib}, \texttt{jupyter} \citep{jupyter}, \texttt{IPython} \citep{ipython}, \texttt{numpy} \citep{numpy}, \texttt{scipy} \citep{scipy}, \texttt{TOPCAT} \citep{topcat}, \texttt{Astropy} \citep{astropy}, \texttt{msaexp} \citep[][]{msaexp}.

\bibliography{MasterBiblio}
\bibliographystyle{apj}

\appendix

\begin{deluxetable}{lc}[h!]
\tabletypesize{\footnotesize}
\tablecaption{Terminal Wind Velocities of LRDs \label{tab:v95}}
\tablehead{
\colhead{Source} & \colhead{$v_{\rm{blue},95\%}$(H$\alpha$)} \\
\colhead{} & \colhead{[km\,s$^{-1}$]} }
\startdata
FRESCO-GN-9771 & $-400^{+6}_{-5}$ \\
FRESCO-GN-15498 & $-239^{+4}_{-6}$ \\
FRESCO-GS-13971 & $-443^{+38}_{-83}$ \\
JADES-GN-68797 & $-1130^{+183}_{-134}$ \\
JADES-GN-38147 & $-411^{+22}_{-26}$ \\
RUBIES-EGS-42046 & $-1041^{+60}_{-120}$ \\
RUBIES-EGS-55604 & $-561^{+19}_{-22}$ \\
RUBIES-EGS-49140 & $-794^{+214}_{-178}$ \\
RUBIES-UDS-182791 & $-569^{+30}_{-26}$ \\
The Cliff & $-145^{+21}_{-111}$ \\
UNCOVER-A2744-45924 & $-389^{+5}_{-8}$
\enddata
\tablecomments{See \S\ref{sec:vesc} and Fig. \ref{fig:v95} for how $v_{\rm{blue, 95\%}}$ is derived, and \citet[][]{Matthee26} for details of the emission line modeling. The sources originate from FRESCO and follow-up H-grating NIRSpec/IFU observations \citep[][]{Oesch23, Matthee24, Torralba25IFU}, JADES \citep[][]{Eisenstein26, CurtisLake26, Scholtz26jades}, RUBIES \citep[][]{degraaff24rubies,Hviding25,degraaff25pop}, and the UNCOVER and ALT surveys \citep[][]{Labbe24, NM24}.}
\end{deluxetable}

\end{document}